\documentclass[12pt]{iopart}

\usepackage{afterpage}
\usepackage{bm}
\usepackage{subeqn}
\usepackage{color}
\usepackage{epsfig,graphics,graphicx,subfigure}
\usepackage[mathcal]{euscript}
\usepackage{float}
\usepackage{hhline}
\usepackage{iopams}

\newcommand{\beqnary}{\begin{eqnarray}}
\newcommand{\eeqnary}{\end{eqnarray}}
\newcommand{\vv}[1]{\vec{#1}}
\newcommand{\ptl}{\partial}
\newcommand{\mm}[1]{\boldsymbol{#1}}
\newcommand{\eqn}[1]{(\ref{#1})}
\newcommand{\twoeqns}[2]{(\ref{#1}) and~(\ref{#2})}

\newcommand{\epsgrid}{\epsilon_g}
\newcommand{\half}{\frac{1}{2}}
\newcommand{\fourpi}{4 \pi}

\begin{document}

\title[The ACC-PIC Method]{An Arbitrary Curvilinear Coordinate Method
  for Particle-In-Cell Modeling}

\author{C.A. Fichtl$^1$, J.M. Finn$^2$, and K.L. Cartwright$^3$}
\address{$^1$ Computational Physics and Methods Group, Los Alamos
  National Laboratory, Los Alamos, NM 87545}
\address{$^2$ Applied Math and Plasma Physics Group, Los Alamos National Laboratory, Los
  Alamos, NM 87545}
\address{$^3$ Electromagnetic Theory Group,
  Sandia National Laboratory, Albuquerque, NM 87185}
\eads{\mailto{cfichtl@lanl.gov}, \mailto{finn@lanl.gov}, \mailto{klcartw@sandia.gov}}
 
\begin{abstract}
A new approach to the kinetic simulation of plasmas in complex
geometries, based on the Particle-in- Cell (PIC) simulation method, is
explored. In the two dimensional (2d) electrostatic version of our method, 
called the Arbitrary Curvilinear Coordinate
PIC (ACC-PIC) method, all essential PIC operations are carried out in
2d on a uniform grid on the unit square logical domain, and mapped to
a nonuniform boundary-fitted grid on the physical domain. As the
resulting logical grid equations of motion are not separable, we have
developed an extension of the semi-implicit Modified Leapfrog (ML)
integration technique to preserve the symplectic nature of the logical
grid particle mover. A generalized, curvilinear coordinate formulation
of Poisson's equations to solve for the electrostatic fields on the
uniform logical grid is also developed. By our formulation, we compute
the plasma charge density on the logical grid based on the particles'
positions on the logical domain. That is, the plasma particles are
weighted to the uniform logical grid and the self-consistent mean
electrostatic fields obtained from the solution of the logical grid
Poisson equation are interpolated to the particle positions on the
logical grid. This process eliminates the complexity associated with
the weighting and interpolation processes on the nonuniform physical
grid and allows us to run the PIC method on arbitrary boundary-fitted
meshes.
\end{abstract}


\maketitle

%\tableofcontents
%%%% TEXT %%%%

%%%%%%%%%%%%%%%%%%%%%%%%%%%%%%%%%%%%%%%%%%%%%%%%%%%%%%%%%%%%%%%%%%%%%%%
%%%%%%%%%%%%%%%%%%%%%%%%%%%%%%%%%%%%%%%%%%%%%%%%%%%%%%%%%%%%%%%%%%%%%%%

\section{INTRODUCTION} \label{sec:intro} 
The standard Particle-in-Cell
(PIC) method utilizes macroparticles to follow the phase space
evolution of weakly coupled (collisionless) plasmas.  In this regime,
PIC can be thought of as an approximate method of integration for the
Vlasov equation~\cite{NicholsonBook}:
\begin{equation} \label{Vlasov} 
\frac{\ptl f_s}{\ptl t} + \vv{v} \cdot \nabla_{\vv{x}} f_s + 
\frac{q_s}{m_s}(\vv{E}+\vv{v} \times \vv{B}) \cdot \nabla_{\vv{v}} f_s = 0,
\end{equation}
where $f_s(\vv{x},\vv{v},t)$ is the time-dependent six-dimensional
phase space distribution function of the plasma particles of species
$s$.  For simplicity, in this work we limit ourselves to the
electrostatic case with no applied background magnetic field.
Applying the method of characteristics~\cite{CarrierPearsonBook} to
(\ref{Vlasov}), it can therefore be shown that the PIC macroparticle
orbits evolve self-consistently in phase-space according to
\begin{equation}\label{eqns of motion}
\begin{array}{rcl}
\dot{\vv{x}} &=& \vv{v} \\
\dot{\vv{v}} &=& -\frac{q_M}{m_M}\nabla \Phi(\vv{x}).
\end{array}
\end{equation}
In \eqn{eqns of motion}, an overdot denotes a time derivative and the
macroparticle charge and mass satisfy $q_M/m_M = q_s/m_s$.  The
mean-field electrostatic potential $\Phi(\vv{x})$ is obtained on a
computational mesh via Poisson's equation
\begin{equation}\label{2d Poisson Eqn}
\nabla^2 \Phi(\vv{x}) = -4\pi \rho(\vv{x}) = - 4\pi \sum_{i = 1}^{N_M} q_M S(\vec{x}-\vec{x_i}).
\end{equation}  
In our notation, $N_M$ is the number of macroparticles (assumed to be
of the same species with equal charge for simplicity) and $\vv{x}_i$
are the particle positions.  $\rho(\vv{x})$ is the charge density
obtained at discrete locations on the mesh from the macroparticles at
$\vv{x}_i$ using interpolation functions, $S(\vec{x}-\vec{x_i})$,
typically chosen to be $B$-splines~\cite{BirdsallLangdonBook}.  These
interpolation functions effectively give the macroparticles a finite
width based upon the grid spacing and lead to cutoff Coulombic
interactions between macroparticles~\cite{Dawson83}.

PIC codes are generally designed using rectangular meshes in Cartesian
geometry, but have been extended to cylindrical and spherical
coordinates.  However, extension to arbitrary grids has proven much
more difficult.  Jones~\cite{Jones87} was among the first to develop a
curvilinear-coordinate PIC method capable of operating on
boundary-conforming grids tailored to accelerator and pulsed-power
applications.  Other codes were soon developed in an effort to model
ion diodes~\cite{Westermann89,Munz99} and microwave
devices~\cite{Eastwood95} more accurately. These early methods involved
generating a nonuniform initial grid based upon the physical
boundaries of the system and running the PIC components on this
physical grid.  There are many benefits to this type of system, such
as having smoothly curved boundaries in contrast to the
``stair-stepped'' boundaries inherent to the rectangular-grid PIC
approach, which occur when a part of the boundary is not aligned with
either coordinate surface. Furthermore, higher grid density can be
placed in areas of interest within the system either statically or by
allowing the grid to adapt
dynamically~\cite{WestermannJCP94,Lapenta07} to the problem by
following a pre-specified control function.  Implemented wisely, such
techniques should allow complex geometries to be simulated at a
fraction of the cost associated with using a uniform grid code, in
which the entire mesh must have the resolution required to resolve the
smallest physical features of the system, and in which the grid does
not conform to the boundary.

Arbitrary grid methods are not without their problems.  Nonuniform
grid cells make it computationally expensive to locate macroparticles
on the grid for the charge accumulation and field interpolation steps
of the PIC method, as well as for enforcing the particle boundary
conditions.  In a standard PIC code with a uniform structured grid,
particle positions on the grid are easily determined.  On a nonuniform
grid, this location must be done iteratively~\cite{SeldnerWestermann},
which increases the computational cost of the method. Furthermore,
interpolation on a nonuniform grid is complicated by
the variations of the grid, which change the shape of the interpolation
functions, often in a non-trivial way~\cite{Westermann92}.  Finally,
as the ratio of the largest to the smallest cell size increases and
the number of particles per cell in the smallest cells becomes small,
the amount of noise near the small cells also increases (assuming the
charge density is roughly constant).  Thus, complex and often
time-consuming gridding strategies~\cite{DelzannoJCP} and/or particle splitting and
merging algorithms~\cite{WelchJCP07} must be implemented to keep the noise within the
system to a minimal threshold value while the structures of interest
within the system are still resolved.

\begin{figure}
\centering
\epsfig{file = 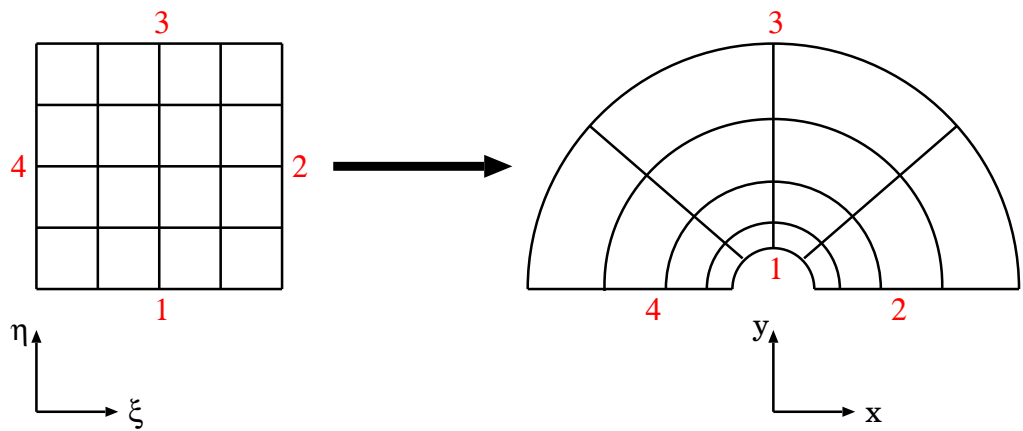,height=2.25in}
\caption{Schematic for mapping from the logical to the physical grid
  for an annulus.  The numbers along the boundaries of the square
  logical domain are used to indicate which edge of the square maps to
  which boundary segment on the physical domain.}
\label{Fig:Grid Mapping}
\end{figure}

We consider the problems associated with these and other existing
adaptive grid PIC approaches to be serious. As such, we have developed
a new nonuniform grid PIC method, which incorporates some of the best
features of several existing methods with a new idea for the
implementation of the PIC method.  Our goal is to design an arbitrary,
curvilinear-coordinate PIC (ACC-PIC) code capable of operating
efficiently and accurately on an arbitrary (but structured) moving
mesh for a boundary of arbitrary geometry.  We construct a logical
(or computational) grid on the unit square and map it to the physical
domain, as illustrated in Figure~\ref{Fig:Grid Mapping}.  We implement the
main components of the PIC method--the charge accumulation, particle
push, field solve, and interpolation--on this logical domain. This 
approach deals with all the problems listed in the previous paragraph 
except the last. The issue of having some cells with much fewer 
macroparticles than others is still a problem requiring attention, 
for example particle splitting and merging.

In this paper, we present results on the development of these methods
into a 2D, electrostatic PIC code.  By implementing the PIC components
on the logical grid, we show that we can eliminate the particle
location and interpolation problems that plagued the earlier boundary
conforming non-uniform grid
methods~\cite{Jones87,Westermann89,Munz99,Eastwood95}.  Particle
locations are easily found on the logical grid using the same
techniques as standard uniform grid PIC codes.  Since the charge
accumulation and field interpolation are both done on the logical
grid, we are again able to use the efficient algorithms that are
utilized in a uniform grid code.  This eliminates the need for
calculating non-standard particle shapes on the physical
grid~\cite{Westermann92}. Furthermore, we develop a Hamiltonian-based,
semi-implicit second-order accurate symplectic logical grid mover and
apply it to the time advance of the particles that includes the
effects of inertial forces.  This mover has the important property of
requiring only a single field solve per timestep.  Since we are moving
particles on a square grid, particle boundary conditions, which may be
difficult to implement on a nonuniform physical grid, are fairly
simple with our approach.  Finally, our method allows us to solve the
electrostatic field equations on a simple square mesh in the logical
domain rather than in the complex physical domain.

The remainder of this paper is structured as follows.  In
Section~\ref{sec:grid generation}, we quickly review the Winslow
method of generating a grid that conforms to a curved boundary, with
finer grid near parts of the boundary with more curvature.  We derive
the particle equations of motion on the logical grid in
Section~\ref{sec:eqns of motion}, and discuss the semi-implicit
modified leapfrog integration technique utilized for the integration
of these equations in Section~\ref{sec:ML}.  Section~\ref{sec:Field
solve} describes the solution of the electrostatic field equations on
the logical grid, as well as details of the charge weighting/field
interpolation schemes utilized.  We perform several standard tests for
the verification of our 2D, electrostatic, nonuniform grid PIC code in
Section~\ref{sec:results}.  Finally, we present our conclusions and
suggestions for future work on the method in
Section~\ref{sec:conclusions}.  A general overview of our differential
geometry notation is given in~\ref{App:DifferentialGeometryNotation}, and a derivation of
the Poisson equation in logical space is given in~\ref{App:
Poisson}.

%%%%%%%%%%%%%%%%%%%%%%%%%%%%%%%%%%%%%%%%%%%%%%%%%%%%%%%%%%%%%%%%%%%%%%%
%%%%%%%%%%%%%%%%%%%%%%%%%%%%%%%%%%%%%%%%%%%%%%%%%%%%%%%%%%%%%%%%%%%%%%%
\section{Logical to Physical Grid Mapping} \label{sec:grid generation}

\subsection{Grid Generation Technique}

For completeness, we include here a discussion of Winslow's Laplace
method~\cite{WinslowJCP}, a simple grid generation technique which
illustrates the logical to physical grid mapping inherent to this
work. In Winslow's method, a set of uncoupled Laplace equations are
solved on a uniform square 2D logical domain, $\xi,\eta \in [0:1]$.
With Winslow's method, the finest gridding is concentrated in the
regions of highest boundary curvature, e.g. around an object at the
center of the domain.  (See Figure~\ref{Fig:Grid Mapping}, in which the
surface labeled ``$1$" might represent a dust grain, ``$3$" represents an
artificial boundary far away, and symmetry is imposed on surfaces
``$2$" and ``$4$."  As discussed in Section~5, this figure can represent
either azimuthal or axial symmetry.)  While this gridding method is
rather primitive in that we have no control over where the grid is
concentrated away from the boundary, we have chosen this method for
its simplicity and suitability for illustrating the feasibility of our
ACC-PIC method.  Winslow's method provides a simple way of generating
an initial boundary conforming grid, to be adapted to the solution at
later timesteps, while still retaining conformation to the boundary
using more sophisticated techniques.

The Laplace equation can be written in any coordinate system, but for
simplicity we have chosen to implement Winslow's method in the
Cartesian coordinate system in physical space.  In the physical
domain, Laplace's equation takes the form
\begin{equation}  \label{lapl phys} 
\nabla_x^2 \xi^\alpha = \frac{\ptl}{\ptl
  x^\beta}\frac{\ptl \xi^\alpha }{\ptl x^\beta} = 0, \quad
\alpha,\beta = 1,\cdots,n.
\end{equation} 
In (\ref{lapl phys}), the logical variables $\xi^\alpha $ are the
dependent variables and the physical variables $x^\beta$ are the
independent variables.  Here and in the rest of this paper we assume
summation over repeated indices.  However, since we would prefer to
solve the Laplace equation on the uniform logical space $(\xi,\eta)$
rather than directly gridding \eqn{lapl phys} on the physical domain,
solving for $\xi(x,y),\eta(x,y)$ and inverting, we must transform the
set of PDE's such that $x^\beta$ are the dependent variables
and $\xi^\alpha $ are the independent variables.  This
transformation has been outlined by Liseikin~\cite{LiseikinBook}; a
detailed derivation can be found in \cite{dissertation}.  After much
manipulation, we write the final system of equations to be solved
as
\begin{equation} \label{winslow equations}
\begin{array}{rcl}
g_{22}\frac{\ptl^2 x}{\ptl \xi^2} - 2g_{12}\frac{\ptl^2
x}{\ptl \xi \ptl \eta} + g_{11}\frac{\ptl^2 x}{\ptl
\eta^2} &=& 0 \\ g_{22}\frac{\ptl^2 y}{\ptl \xi^2} -
2g_{12}\frac{\ptl^2 y}{\ptl \xi \ptl \eta} +
g_{11}\frac{\ptl^2 y}{\ptl \eta^2} &=& 0,
\end{array}
\end{equation}
where 
\begin{equation} \label{cov metric}
g_{\mu \nu}(\vv{\xi}\,) \equiv \frac{\ptl x^\gamma}{\ptl \xi^\mu }\frac{\ptl
x^\gamma}{\ptl \xi^\nu},\quad \mu,\nu,\gamma = 1,\cdots,n
\end{equation}
is the covariant metric tensor (see Appendix A). While
(\ref{winslow equations})(a) and (b) appear to be the same
equation for $x(\xi,\eta)$ and for $y(\xi,\eta)$, they actually
possess opposite boundary conditions along each segment of the grid
boundary: For any segment of the boundary, we have Neumann boundary
conditions for $\xi$ and Dirichlet boundary conditions for $\eta$ or
vice-versa, both expressed in terms of $x(\xi,\eta)$ and
$y(\xi,\eta)$.  Furthermore, from \eqn{lapl phys} and these boundary
conditions, we know that $\xi(x,y)$ and $\eta(x,y)$ are conjugate
harmonic functions.  Thus, Winslow's method forces the grid lines to
be orthogonal (i.e. $g_{12} = 0$) in the physical space.

Since the covariant terms $g_{\alpha\beta}$ depend on
$\frac{\partial x_{\alpha}}{\partial \xi_{\beta}}$ and the solution we seek is 
$x(\xi,\eta),y(\xi,\eta)$, the
system of equations is nonlinear.  (\ref{winslow
equations}) are therefore discretized using a second-order accurate
finite-difference method and, because of their non-linear form, are
iteratively solved using an inexact Newton-Krylov
solver with the Generalized Minimal Residual Method (GMRES)~\cite{KelleyBook,Dembo}.

\begin{figure}
\centering \subfigure[]{
  \includegraphics[width=2.25in]{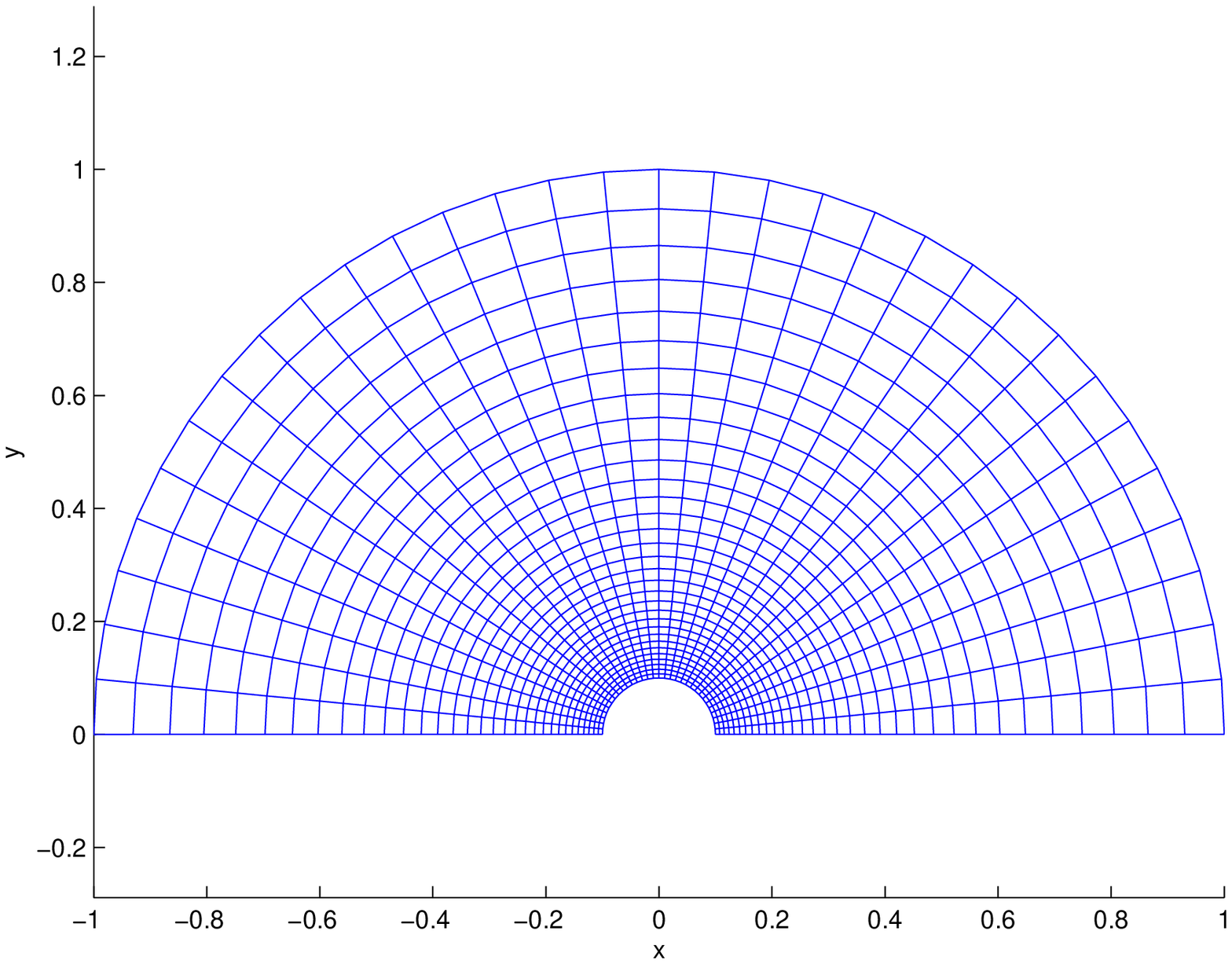}}
  \subfigure[]{
\includegraphics[width=2.25in]{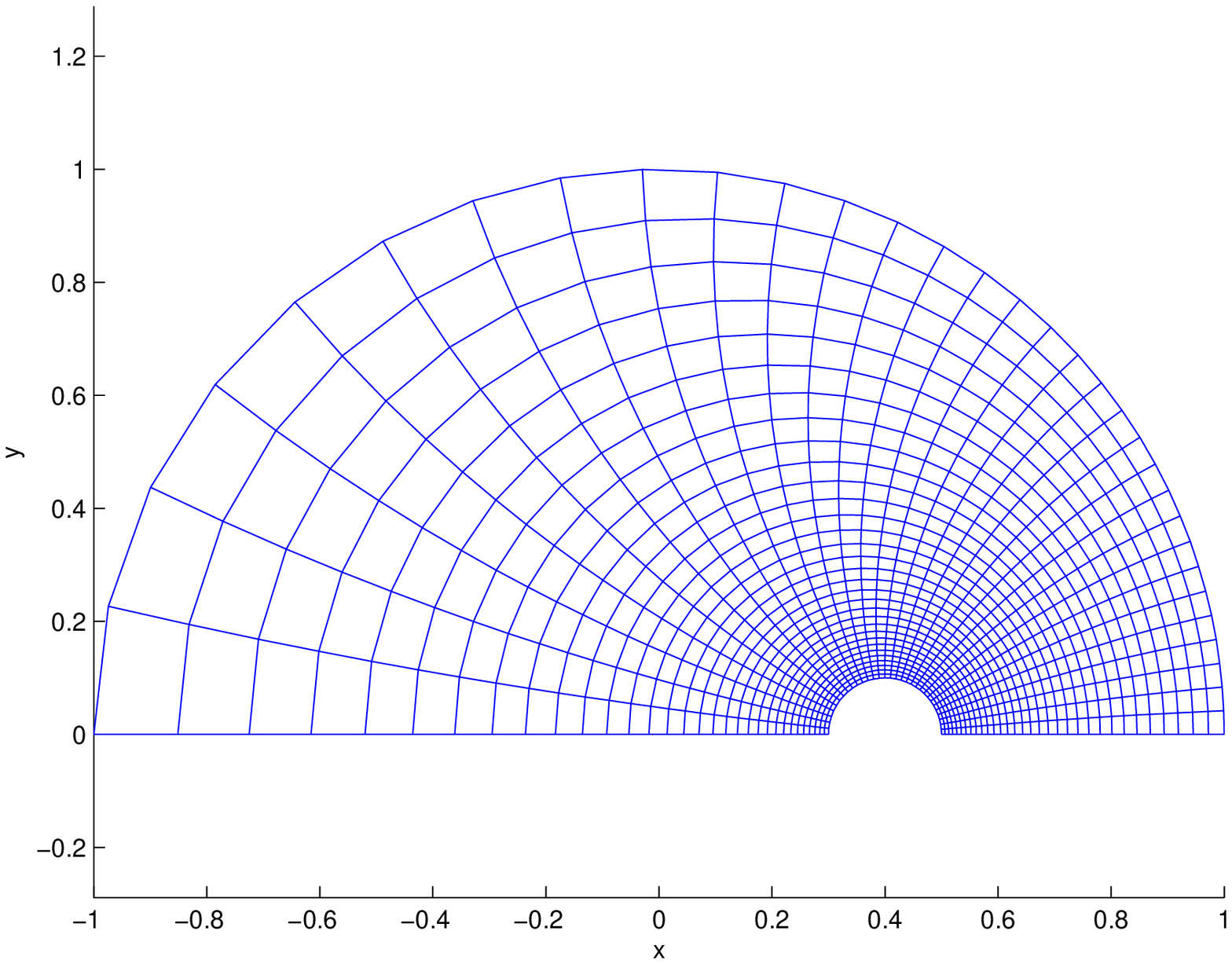}}
  \subfigure[]{
  \includegraphics[width=2.25in]{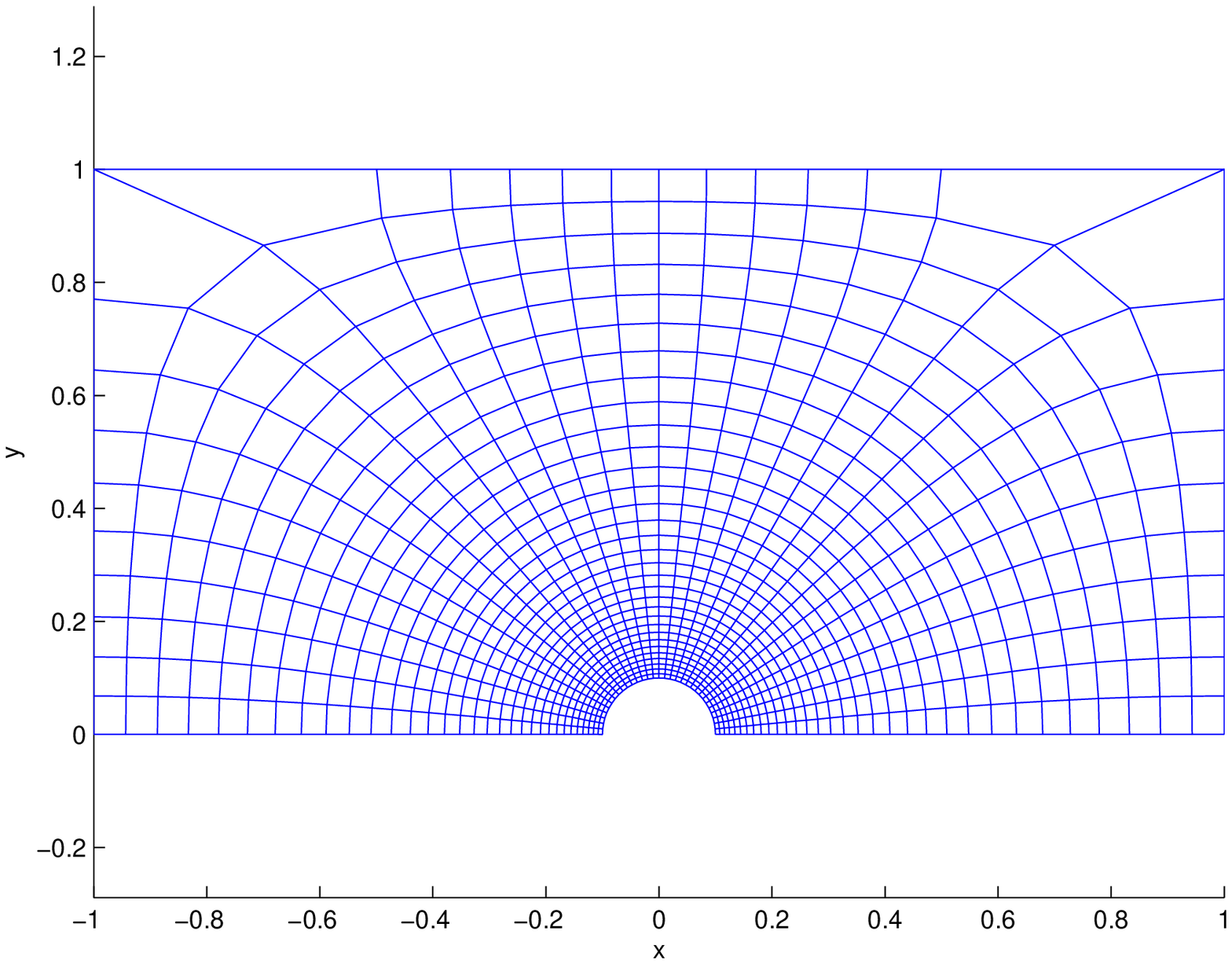}}
  \subfigure[]{
  \includegraphics[width=2.25in]{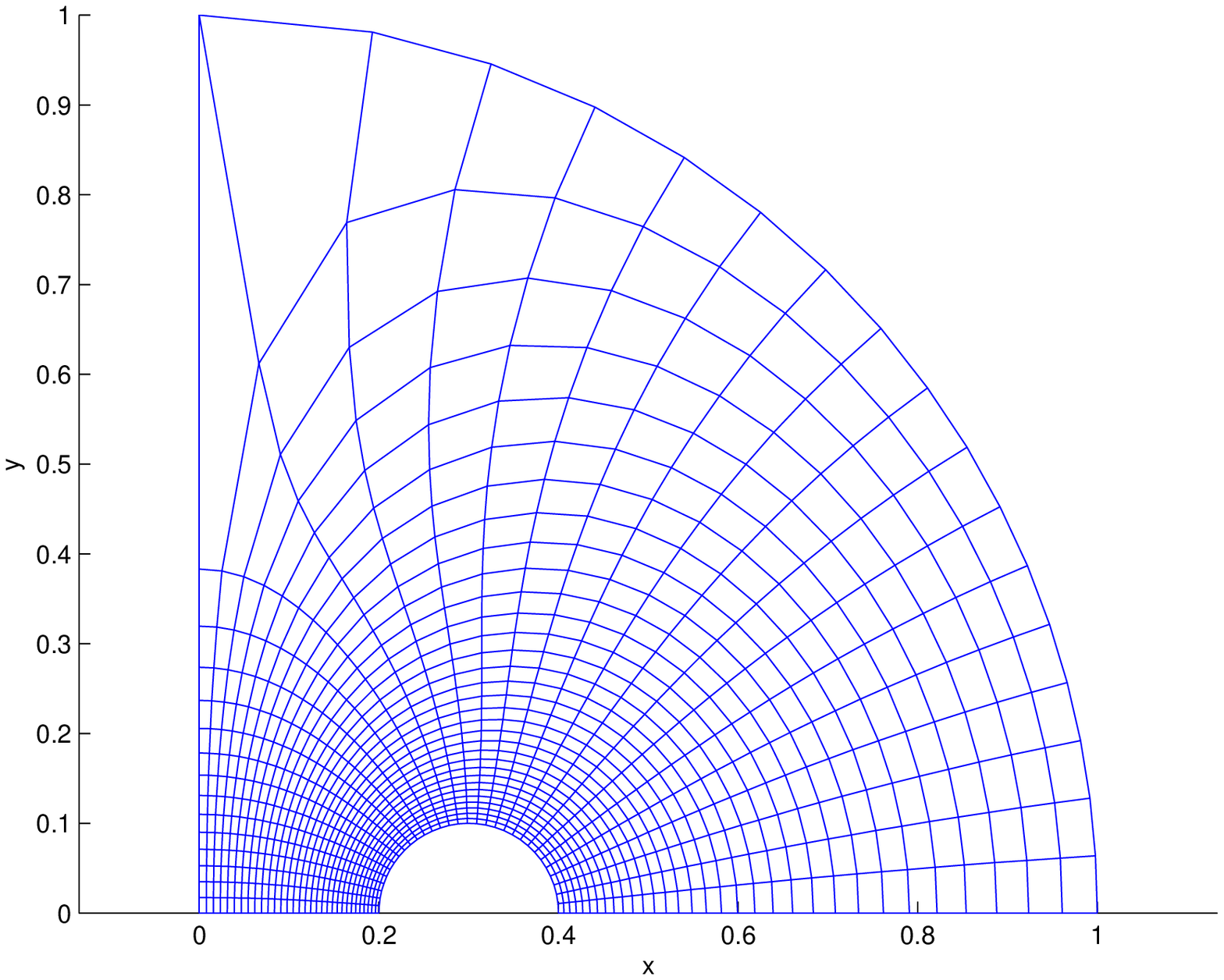}}
\caption{Grids generated using Winslow's method with a circular
  object along the inner physical grid boundary, symmetry along the bottom 
  segments (and on the left in (d)), and an
artificial boundary far away at top.  Here we have used
  a $32\times32$ grid in order to show the mapping more clearly.
  The appearance of nonorthogonal grid lines is due to the straight
  lines used by the plotting tool.}
\label{Fig:Grids}
\end{figure}

Winslow grids can be specified for a wide array of physical domains by
specifying boundary conditions on boundary segments, as illustrated in
Figure~\ref{Fig:Grid Mapping}.  The grids presented in
Figure~\ref{Fig:Grids} are of some of the physical grids with circular
objects along the inner boundary which we have generated using
Winslow's method.  As we discuss in Section~5, this circular segment
can represent a cylindrical or spherical surface.  We have also
generated grids with elliptically shaped objects, and are capable of
generating objects of fairly general shape and size on any boundary
segment~\cite{dissertation}.

%%%%%%%%%%%%%%%%%%%%%%%%%%%%%%%%%%%%%%%%%%%%%%%%%%%%%%%%%%%%%%%%%%%%%%%
%%%%%%%%%%%%%%%%%%%%%%%%%%%%%%%%%%%%%%%%%%%%%%%%%%%%%%%%%%%%%%%%%%%%%%%

\section{DEVELOPMENT OF LOGICAL GRID EQUATIONS OF MOTION} \label{sec:eqns of motion}

The most obvious method for implementing our mover in logical space
would seem to be by simply converting the Newton-Lorentz equations of
motion from the physical to the logical grid.  Specializing to 1d for
simplicity and defining $u = \dot{\xi} = v/J$, we recast
(\ref{eqns of motion}) as
\begin{equation}\label{NL trans to log EOM 1d}
\begin{array}{rcl}
\dot{\xi} &=& f(u) = u \\ 
\dot{u} &=& g(\xi,u) = -\frac{J'u^2}{J} - \frac{q}{m}\frac{1}{J^2}\Phi',
\end{array}
\end{equation}
where $J \equiv \frac{d x}{d\xi}$ in 1d and the prime symbol
represents a derivative with respect to the logical coordinate $\xi$.
For compactness we have dropped the subscripts designating
macroparticle quantities here and in the rest of this paper.  Taking
the formal divergence of (\ref{NL trans to log EOM 1d}), we have
\begin{equation} \label{div of 1d NL trans to log EOM 1d}
\frac{\ptl f}{\ptl \xi}\bigg{|}_u+\frac{\ptl
g}{\ptl u}\bigg{|}_\xi = -\frac{2 J' u}{J} \neq 0.
\end{equation}
The Newton-Lorentz equations of motion are therefore not divergence-free 
in these variables.  As such, a time integration of these
equations of motion with a standard ``naive'' leapfrog (LF) integrator,
\begin{equation} \label{naive LF}
\begin{array}{rcl}
\xi' &=& \xi + \Delta t \, f(u) \\
u' &=& u + \Delta t \, g(\xi',u),
\end{array}
\end{equation}
will not preserve phase space area, and the particle orbits will
typically spiral inward or outward with time~\cite{dissertation}.
  
\subsection{Hamiltonian Approach to Logical Grid Equations of Motion}

Clearly, since the system of equations in (\ref{NL trans to log EOM
  1d}) are not in a canonical Hamiltonian formulation, transformation
of the Newton-Lorentz equations of motion into logical coordinates
leads to a lack of $(\xi,u)$ phase space area conservation.  We
therefore construct logical grid particle equations of motion based on
Hamilton's equations by using a canonical transformation
$(\vv{x},\vv{p}) \rightarrow (\vv{\xi},\vv{P})$, where
$p^\alpha=mv^\alpha$ is the physical space momentum and $P^\alpha$ is
the logical space momentum.  This transformation is constructed via an
$F_2(\vv{x},\vv{P},t)$ generating function~\cite{GoldsteinBook},
thereby assuring Hamiltonian equations of motion on the logical grid.
Under the canonical transformation we have
\begin{equation}\label{gen function}
\begin{array}{rcl}
p^\alpha &=& \frac{\ptl }{\ptl x^\alpha} F_2(\vv{x},\vv{P},t)\\
\xi^\alpha &=& \frac{\ptl }{\ptl P^\alpha}F_2(\vv{x},\vv{P},t).
\end{array}
\end{equation}
We have chosen the $F_2$ generating function such that 
$\vv{x}$ and $\vv{P}$ are considered
independent variables.  The transformed Hamiltonian $K$ is given by
\begin{equation}\label{contact trans}
K = H + \frac{\ptl F_2}{\ptl t}.
\end{equation}
A time-adaptive grid is beyond the scope of this paper; we therefore drop
the second term of (\ref{contact trans}).  Specializing
to the contact transformation $\vv{\xi} = \vv{\xi}(\vv{x})$, we
write the $F_2(\vv{x},\vv{P})$ generating function as
\begin{equation}\label{contact trans final}
F_2(\vv{x},\vv{P}) = \xi^\beta(\vv{x}) P^\beta,
\end{equation}
such that by (\ref{gen function}), we have $p^\alpha = k^{\beta\alpha}
P^{\beta}$ and $\xi^\alpha = \xi^{\alpha}(\vv{x})$.  Here $
k^{\beta\alpha} $ is the inverse Jacobi matrix as defined in Appendix
A.

In the absence of a static background magnetic field $\vv{B}$, the
physical grid Hamiltonian is given in general form by
\begin{equation}\label{Physical Hamiltonian}
H = \sum_i \frac{p_i p_i}{2m} + q\Phi(\vv{x}),
\end{equation}
where the sum is over three rectangular components in physical
space. Notice that (\ref{Physical Hamiltonian}) is fully
separable, $H = T(\vv{p}) +V(\vv{x})$.  In this work, we consider two
distinct 2D cases, with azimuthal ($z$) symmetry and axial ($\phi$)
symmetry.  In the former, the grid segment labeled ``$1$" along the
curved grid segment in Figure~\ref{Fig:Grid Mapping} represents, for
example, a cylindrical probe.  For the axisymmetric case, this segment
would represent a spherical probe.  We stress here that both cases
lead to separable Hamiltonians in the physical domain.

(\ref{Physical Hamiltonian}) can then be used in conjunction
with (~\ref{contact trans}) to construct the logical grid Hamiltonian:
\begin{equation} \label{Hamiltonian}\nonumber 
K = \frac{1}{2m}\left(g^{\beta \gamma}P^\beta P^\gamma \right) +
V(\vv{\xi} ),
\end{equation}
where $ g^{\beta \gamma} $ is the contravariant metric tensor as
defined in \eqn{contra metric}.  Again, since
$g^{\beta\gamma}=g^{\beta\gamma}(\xi,\eta)$, we must represent
$g^{\beta\gamma}$ 
in terms of the covariant metric tensor as in
(\ref{App:cov to cont}).  Note that on the logical grid the
transformed Hamiltonian can be written $K = T(\vv{\xi},\vv{P}) +
V(\vv{\xi})$, meaning that we have transformed the separable physical
grid Hamitonian to an equivalent, but non-separable system.  Applying
Hamilton's equations, $\dot{\xi}^\alpha \equiv \frac{\ptl K}{\ptl
  P^\alpha }$ and $\dot{P}^\alpha \equiv -\frac{\ptl K}{\ptl
  \xi^\alpha }$ to (\ref{Hamiltonian}) gives the logical grid
equations of motion:
\begin{equation}\label{Hamiltonian EOM}
\begin{array}{rcl}
\dot{\xi}^\mu &=& \frac{g^{\gamma \mu}P^\gamma}{m}=U^{\mu}(\vv{\xi},\vv{P})\\
\dot{P}^\mu &=& -\frac{1}{2m}\frac{\ptl g^{\beta \gamma}}{\ptl
\xi^\mu}P^\beta P^\gamma - \frac{\ptl V}{\ptl \xi^\mu}=W^{\mu}(\vv{\xi},\vv{P}).
\end{array}
\end{equation}
We note here that the term quadratic in $\vv{P}$ represents the
inertial force in these coordinates, and that both $\dot{\xi}^\mu$ and
$\dot{P}^\mu$ are functions of $\xi^\alpha$ and $P^\alpha$.  Since
these equations are obtained from a Hamiltonian in canonical variables, 
the divergence for
each degree of freedom $\frac{\partial U^{\mu}}{\partial \xi^{\mu}} +
\frac{\partial W^{\mu}}{\partial P^{\mu}}$ (no sum) is zero.
%%%%%%%%%%%%%%%%%%%%%%%%%%%%%%%%%%%%%%%%%%%%%%%%%%%%%%%%%%%%%%%%%%%%%%%
%%%%%%%%%%%%%%%%%%%%%%%%%%%%%%%%%%%%%%%%%%%%%%%%%%%%%%%%%%%%%%%%%%%%%%%

\section{PARTICLE PUSH: MODIFIED LEAPFROG INTEGRATOR} \label{sec:ML}

Whereas the physical space Hamiltonian leads to separable equations of
motion which can be integrated by standard LF integration techniques,
in the logical space this is no longer true.  Since (\ref{Hamiltonian
  EOM}) is not a separable system of equations, integration with the
LF method will not conserve phase space area.

As such, we have chosen to implement an extension of the semi-implicit
modified leapfrog (ML) integrator originally developed by
Finn and Chac\'on~\cite{FinnChacon05} for integrating 2D solenoidal
flows in fluid dynamics and magnetic field lines in MHD codes. (Because 
the dependence of the Hamiltonian on the canonical momentum is given analytically, 
the exactly divergence free interpolations used in~\cite{FinnChacon05} 
are not necessary.)
Rewriting (\ref{Hamiltonian EOM}) as $\dot{\vv{P}} = \vv{W}$ and
$\dot{\vv{\xi}} = \vv{U}$, respectively, and denoting explicit
(implicit) updates with a superscript $e$ ($i$), the ML integrator can
be written as $\vv{M}_{\Delta t} = \vv{P}^e_{\Delta t} \circ
\vv{\xi}^i_{\Delta t}$, where
\begin{equation} \label{modlf general}
\vv{\xi}^i_{\Delta t}:\left\{ \begin{array}{c} \vv{\xi}_1 = \vv{\xi} +
\Delta t \, \vv{U}(\vv{\xi}_1,\vv{P}) \\ \vv{P}_1 = \vv{P}
\end{array}\right., \quad \vv{P}^e_{\Delta t}:\left\{ \begin{array}{c} \vv{\xi}' =
\vv{\xi}_1 \\ \vv{P}' = \vv{P}_1+ \Delta t \,\vv{V}(\vv{\xi}_1,\vv{P}_1)
\end{array}\right. .
\end{equation}
Combining, we have
\begin{equation} \label{modlf comb}
\begin{array}{rcl}
\vv{\xi}' &=& \vv{\xi} + \Delta t \, \vv{U}(\vv{\xi}',\vv{P}) \\
\vv{P}' &=& \vv{P} + \Delta t \, \vv{V}(\vv{\xi}',\vv{P}).
\end{array}
\end{equation}
The map $\xi^i_{\Delta t}$ is implicit and must be done by means of
Newton or Picard iterations.  The map $P^e_{\Delta t}$ is explicit,
and can therefore be applied directly.  It can be shown that, unlike
the standard LF scheme commonly utilized in PIC codes, this
non-time-centered formulation of the ML scheme results in only
first-order accuracy in $\Delta t$. To achieve second-order accuracy
in time, we simply symmetrize the ML scheme by composing
$\vv{\xi}^e_{\Delta t/2} \circ \vv{P}^i_{\Delta t/2} \circ
\vv{P}^e_{\Delta t/2} \circ \vv{\xi}^i_{\Delta t/2}$. As shown
in~\cite{FinnChacon05}, the implicit-followed-by-explicit ordering in
each pair of mappings retains the area preserving nature of the
integrator in a 2D phase space (one degree of freedom.) This
integrator is seen to be symplectic for arbitrary degrees of freedom
because it can be derived from a generating
function~\cite{GoldsteinBook}
\begin{equation} \label{F3}
F_3(\vv{\xi}',\vv{P})=-\xi'^{\alpha}P^{\alpha}+\Delta t K(\vv{\xi}',\vv{P}),
\end{equation}
with
\begin{equation}\label{F3 gen function}
\begin{array}{rcl}
P'^\alpha &=& -\frac{\ptl }{\ptl \xi'^\alpha} F_3(\vv{\xi}',\vv{P})\\
\xi^\alpha &=& -\frac{\ptl }{\ptl P^\alpha}F_3(\vv{\xi}',\vv{P}).
\end{array}
\end{equation}
The second half of the ML update ($\vv{P}^e_{\Delta t/2} \circ
\vv{\xi}^i_{\Delta t/2}$) is described by an $F_2(\vv{\xi},\vv{P}')$
generating function. The alternation of the steps in $\vv{\xi}$ and
$\vv{P}$ gives second-order accuracy in $\Delta t$.  The logical flow
of the ML integrator as implemented on (\ref{Hamiltonian EOM}) is
\begin{equation} \nonumber \label{mover flow}
\vv{\xi}^i_{\Delta t/2} \rightarrow \vv{P}^e_{\Delta t/2}
\rightarrow \vv{P}^i_{\Delta t/2} \rightarrow \vv{\xi}^e_{\Delta t/2}.
\end{equation}
We note here that the charge density is accumulated on the grid and
the mean field solve is performed after the implicit position
update step of the symmetrized ML mover.  While this mover requires us to
pass through the particle array twice per timestep, it allows us to
accumulate the charge density and solve for the fields only \textit{once} per
timestep.

With the exception of the case in Sec.~7.4, the boundary
conditions we use are periodic, and we therefore do not need to
describe particle reflection in the logical space. Reflecting
conditions on particles can be implemented with some care in logical
coordinates.

%%%%%%%%%%%%%%%%%%%%%%%%%%%%%%%%%%%%%%%%%%%%%%%%%%%%%%%%%%%%%%%%%%%%%%
%%%%%%%%%%%%%%%%%%%%%%%%%%%%%%%%%%%%%%%%%%%%%%%%%%%%%%%%%%%%%%%%%%%%%%

\section{FIELD SOLVER: GENERALIZED POISSON EQUATION} \label{sec:Field
solve} 

In order to solve the electrostatic field equation on the
logical grid, we first write the Poisson equation on the physical grid
\begin{equation} \label{poisson with f} 
\frac{1}{f}\nabla \cdot f\nabla \Phi = \frac{1}{f}\frac{\ptl }{\ptl
x^\alpha } \cdot f\frac{\ptl \Phi}{\ptl x^\alpha } =
-\fourpi \rho^x,
\end{equation} 
where $\rho^x$ is the physical charge density.  Here, $f$ is
a geometry factor. For azimuthal symmetry, $f=1$ and the
Poisson equation takes the usual form 
\begin{subequations}
\begin{equation}\label{Poisson azi}
\nabla^2 \Phi = -\fourpi \rho^x. 
\end{equation}
For axisymmetry, we have $f=r$ and the Poisson equation takes
the form
\begin{equation}\label{Poisson axi}
\nabla \cdot \left(r \nabla \Phi \right) = -\fourpi r \rho^x.
\end{equation}
\end{subequations}
The generalized curvilinear coordinate formulation of Poisson's
equation on the logical grid takes the form
\begin{equation}\label{curv coords poisson} 
\frac{1}{fJ} \frac{\ptl}{\ptl \xi^\alpha } \left(fJ g^{\alpha
\beta}\frac{\ptl \Phi}{\ptl \xi^\beta} \right) =-
\fourpi\rho^x,  
\end{equation} 
as derived in Appendix B. The logical density $\rho^{\xi}$
is equal to $J\rho^x$, leading to
\begin{equation}\label{curv coords poisson-2} 
\frac{\ptl}{\ptl \xi^\alpha } \left(fJ g^{\alpha
\beta}\frac{\ptl \Phi}{\ptl \xi^\beta} \right) =-
\fourpi f \rho^{\xi}.  
\end{equation} 
In this form, we accumulate the logical density $\rho^{\xi}$ and solve
the Poisson equation by conservative differencing, as described in the
next section.  This logical grid solver removes the necessity of
writing and maintaining multiple complex-geometry Poisson solvers for
the various coordinate systems we wish to model.

%%%%%%%%%%%%%%%%%%%%%%%%%%%%%%%%%%%%%%%%%%%%%%%%%%%%%%%%%%%%%%%%%%%%%%%
%%%%%%%%%%%%%%%%%%%%%%%%%%%%%%%%%%%%%%%%%%%%%%%%%%%%%%%%%%%%%%%%%%%%%%%

\section{NUMERICAL IMPLEMENTATION} \label{sec:implementation}

For our 2D code, we have a wide range of grid choices on which to
perform validation tests for our method in this and the next section.
For the sake of brevity, we outline only two here, both of which have
been designed such that analytical expressions for the metric tensors
are easily obtained.  In addition to the annulus grid generated
numerically by the methods described in Section~\ref{sec:grid
  generation}, we define a doubly periodic, nonuniform, orthogonal
grid using:
\begin{subequations}\label{2d grids}
\begin{equation}\label{periodic sin grid}
\begin{array}{rcl}
x &=& x_\mathrm{min}+(x_\mathrm{max}-x_\mathrm{min})(\xi + 
\epsilon_g \sin{2 \pi \xi}) \\
y &=& y_\mathrm{min}+(y_\mathrm{max}-y_\mathrm{min})(\eta + 
\epsilon_g \sin{2 \pi \eta}).
\end{array}
\end{equation}
Here $x_\mathrm{min}$, $x_\mathrm{max}$, $y_\mathrm{min}$, and
$y_\mathrm{max}$ are constants to scale boundaries of the physical
grid to form a rectangle of arbitrary size. Here, $\epsgrid$ is the
nonuniformity parameter which controls the amount of nonuniformity of
the grids.  A doubly-periodic grid has been chosen for the implementation of
the periodic field and particle boundary conditions utilized for
the tests in the Section~\ref{sec:results}.  We have also designed a
doubly-periodic, nonorthogonal grid given by
\begin{equation}\label{nonorthogonal grid}
\begin{array}{rcl}
  x &=& x_\mathrm{min}+(x_\mathrm{max}-x_\mathrm{min})(\xi + 
\epsilon_g \sin{2 \pi \xi} \sin{2 \pi \eta})\\
  y &=& y_\mathrm{min}+(y_\mathrm{max}-y_\mathrm{min})(\eta + 
\epsilon_g \sin{2 \pi \xi} \sin{2 \pi \eta})
\end{array}
\end{equation}
\end{subequations}
to test the effects of non-zero cross terms ($g^{12}$) in our code,
both in the field solver and the particle mover.  For simplicity, we
have constrained the grid nonuniformity parameter
$\epsilon_g$ in \twoeqns{periodic sin grid}{nonorthogonal
  grid} to be the same in each dimension.  Notice that both
\twoeqns{periodic sin grid}{nonorthogonal grid} require that $\epsgrid
< \frac{1}{2\pi}$ so that the grid does not fold.

\subsection{Discretization of Particle Equations}

In two spatial dimensions, the logical grid particle equations of
motion ((\ref{Hamiltonian EOM})) can be written as
\begin{subequations}\label{2d Ham Log Eqns of Motion}
\begin{equation}\label{Ham 2d pos update}
\begin{array}{rcl}
\vphantom{\Biggl (}\dot{\xi} &=& \frac{1}{m}\left(g^{11}P_{\xi} +
g^{12}P_{\eta}\right) \\
\vphantom{\Bigl (}\dot{\eta} &=& \frac{1}{m}\left(g^{12}P_{\xi} + 
g^{22}P_{\eta} \right),
\end{array}
\end{equation}
and
\begin{equation}\label{Ham 2d mom update}
\begin{array}{rcl}
\vphantom{\Biggl (}\dot{P}_{\xi} &=& \frac{-1}{2m}\left(P_{\xi}^2
\frac{\partial g^{11}}{\partial \xi} + 2
P_{\xi}P_{\eta}\frac{\partial g^{12}}{\partial
\xi}+P_{\eta}^2\frac{\partial g^{22}}{\partial
\xi}\right) - \frac{\partial V(\vv{\xi}\,)}{\partial \xi} \\ 
\vphantom{\Biggl (}\dot{P}_{\eta} &=&
\frac{-1}{2m}\left(P_{\xi}^2 \frac{\partial g^{11}}{\partial
\eta} + 2 P_{\xi}P_{\eta}\frac{\partial g^{12}}{\partial
\eta} + P_{\eta}^2\frac{\partial g^{22}}{\partial
\eta}\right) - \frac{\partial V(\vv{\xi}\,)}{\partial \eta}.
\end{array}
\end{equation}
\end{subequations}
Note that we have written (\ref{2d Ham Log Eqns of Motion}) in
terms of the contravariant metric tensor, $g^{\mu \nu} = g^{\mu
\nu}(\vv{x}\,)$, which is easily obtained as the inverse of the
covariant metric tensor, $g_{\mu \nu}(\vv{\xi}\,)$ as in \eqn{App:cov
to cont}.

The third momentum component is generated at $t = 0$ and held
as a constant as the simulation progresses.  This term contributes to
the simulation through its inclusion in the effective potential term,
$V(\vv{\xi}\,)$.  In azimuthal symmetry,  $(x_1,x_2)=(x,y)$ and the 
effective potential is $V_\mathrm{azi}(\vv{\xi}\,) =
\frac{p_z^2}{2m} + q\Phi(\vv{\xi}\,)$, such that the
ignorable-direction momentum does not contribute to the momentum
update equations.  However, for an axisymmetric problem, 
 $(x_1,x_2)=(r,z)$ and the effective potential is
$V_\mathrm{axi}(\vv{\xi}\,) = \frac{p_\phi^2}{2 m r(\vv{\xi}\,)^2} +
q\Phi(\vv{\xi}\,)$, such that its derivative is
\begin{equation}\label{axi effective pot deriv}
\begin{array}{rcl}
  \vphantom{\Bigl (}\frac{\partial V(\vv{\xi}\,)}{\partial \xi} &=&
  -\frac{j_{11} p_\phi^2}{m r^3} - q E_\xi\\ \vphantom{\Biggl
  (}\frac{\partial V(\vv{\xi}\,)}{\partial \eta} &=&-\frac{j_{12}
  p_\phi^2}{m r^3} - q E_\eta,
\end{array}
\end{equation}
where $(x_1,x_2)=(r,z)$ implies $j_{11}=\partial r / \partial \xi$ and $j_{12}=\partial r / \partial
\eta$. Thus, in the axisymmetric case we must also interpolate the
$j_{11}$ and $j_{12}$ components of the Jacobi matrix and the
$r$-coordinate of the particle's physical space position to the
particle position on the logical grid.

\subsection{Conservative discretization of Poisson equation}

In two dimensions the logical grid Poisson equation, \eqn{curv coords
  poisson-2}, takes the form
\begin{equation}\label{2d Logical Grid Poisson}
\frac{\partial}{\partial
\xi}\left(D^{11}\frac{\partial
\Phi}{\partial \xi} + D^{12}\frac{\partial \Phi}{\partial
\eta}\right) + \frac{\partial}{\partial
\eta}\left(D^{12}\frac{\partial
\Phi}{\partial \xi} + D^{22}\frac{\partial \Phi}{\partial
\eta}\right) = -\fourpi \mathit{f}\rho^\xi,
\end{equation}
where we have defined $D^{\mu \nu} \equiv \mathit{f}Jg^{\mu \nu}$,
again written in terms of the contravariant metric tensor $g^{\mu
  \nu}(\vv{x})$.  \eqn{2d Logical Grid Poisson} is comprised of a set
of co-directed derivatives proportional to the diagnonal metric tensor
components ($D^{11}$ and $D^{22}$) and a set of cross-directed
derivatives for the off-diagonal metric tensor components ($D^{12}$),
i.e.  $\nabla^2 \Phi = (\nabla^2\Phi)^\mathrm{co} +
(\nabla^2\Phi)^\mathrm{cross}$, the latter of which are nonzero for
non-orthogonal coordinates ($g^{12} \neq 0$).  We have defined the
coordinates $\xi,\eta$ at vertices $i,j$, thus $\Phi$ and $\rho$ are
naturally defined at cell centers $i+1/2,j+1/2$.  The co- and
cross-directed terms are discretized using appropriate
centered-difference schemes, leading to the final discretized form of
\eqn{2d Logical Grid Poisson}:
%\begin{scriptsize}
\begin{equation}\label{discrete 2d Logical Grid Poisson}\nonumber
\begin{array}{c}
 \hspace{-120mm} -\fourpi \mathit{f}\rho^\xi_{i+\half,j+\half} =\\
  \frac{D^{11}_{i+1,j+\half}\left(\Phi_{i+\frac{3}{2},j+\half}-\Phi_{i+\half,j+\half}
  \right) -
  D^{11}_{i,j+\half}\left(\Phi_{i+\half,j+\half}-\Phi_{i-\half,j+\half}
  \right)}{\Delta \xi^2}+\\ \nonumber
  
  \frac{D^{22}_{i+\half,j+1}\left(\Phi_{i+\frac{1}{2},j+\frac{3}{2}}-\Phi_{i+\half,j+\half}\right)-
  D^{22}_{i+\half,j}\left(\Phi_{i+\half,j+\half}-\Phi_{i+\half,j-\half}
  \right)}{\Delta \eta^2}+\\ \nonumber 
  
  \frac{1}{4\,\Delta
  \xi\,\Delta \eta} \left[D^{12}_{i+1,j+1}
  \left(\Phi_{i+\frac{3}{2},j+\frac{3}{2}}-\Phi_{i+\half,j+\half}\right)
  -
  D^{12}_{i,j}\left(\Phi_{i+\half,j+\half}-\Phi_{i-\half,j-\half}\right)
  + \right.\\ 
  
  \left.
  D^{12}_{i+1,j}\left(\Phi_{i+\half,j+\half}-\Phi_{i+\frac{3}{2},j-\half}\right)
  +
  D^{12}_{i,j+1}\left(\Phi_{i-\half,j+\frac{3}{2}}-\Phi_{i+\half,j+\half}\right)
  \right].
\end{array}
\end{equation}
%\end{scriptsize}

Note that, for a uniform grid with 
$x,y \in [0:1]$, the $D^{11}$ and
$D^{22}$ terms become unity (assuming the geometry factor $\mathit{f}
= 1$) and $D^{12} = 0$, thus \eqn{discrete 2d Logical Grid Poisson}
reduces to the standard $5$-point discretization commonly used in
Cartesian PIC codes.

We note that this same discretization of \eqn{2d Logical Grid Poisson}
can also be obtained by the minimization of the variational
principle~\cite{LuisPrivateConv10}
\begin{equation}\label{Poisson Variational Princ}
W = \int \left[\frac{\left|\nabla \Phi \right|^2}{2} - \rho^x \Phi\right] dV,
\end{equation}
discretized on the logical grid.  The variational principle approach
guarantees that, for properly applied boundary conditions, the matrix
formed by the application of the discrete form of the $\nabla^2$
operator is symmetric (for appropriate boundary conditions) and
negative definite.  This property is important since it permits the
use a fast conjugate gradient (CG) solver. Symmetry is important
because CG requires a symmetric, positive (or negative) definite
matrix to converge to the correct solution, but for a
non-symmetric matrix it may converge to the wrong answer.

\subsubsection{Validation of Poisson Solver using the Method of Manufactured Solutions}

The Method of Manufactured Solutions (MMS)~\cite{MMS02} technique is
utilized to validate our Poisson solver.  MMS is a simple, yet
powerful tool to construct solutions to PDE problems. By obtaining
analytic solutions to problems, we are able to check that the error in
the numerical solution converges to the analytic solution with
second-order accuracy in grid spacing. MMS is based upon choosing a potential $\Phi$ which
satisfies a given set of boundary conditions \textit{a priori}, then
taking the required derivatives to find the source term, i.e.~the density. 
To validate our solution of Poisson's
equation, we simply choose a potential that satisfies a chosen set of
boundary conditions, calculate the charge density, and use that
density in our solver.  Upon convergence of the solver, the numerical
solution and the exact solution are compared, and the error between
the two is calculated using the $L_2$-norm:
\begin{equation}\label{L_2 Norm}
||\Delta \Phi||_2 = \sqrt{\frac{\sum_{i=1}^{N_\xi} \left(\Phi^{\mathrm{num}}_i -
\Phi^{\mathrm{MMS}}_i\right)^2}{N_\xi}}.
\end{equation}
To ensure the accuracy of our method, we 
have set up MMS tests appropriate for both the analytically given grids of (25)
as well as for the case in which an
annular grid is numerically generated using the techniques of
Section~\ref{sec:grid generation}.  

\begin{table}[t!]
\centering
\small
\centering
\begin{tabular}{|c|c|c|c|c|c|}\hline
  $\mm{N_\xi}$ & $\mm{ \epsilon_{g} = 0}$ & $
  \mm{\epsilon_{g} = 0.025}$ & $\mm{\epsilon_{g} =
    0.05}$& $ \mm{\epsilon_{g} = 0.075}$ & $\mm{
    \epsilon_{g} = 0.15}$\\\hline\hline 
16 & 1.72090E-3 & 8.29163E-4 & 6.88949E-4 & 2.42544E-3 & 9.06392E-3  \\\hline 
32 & 4.14875E-4 & 1.97193E-4 & 1.69813E-4 & 5.95391E-4 & 2.21714E-3  \\\hline 
64 & 1.02009E-4 & 4.83358E-5 & 4.20043E-5 & 1.47031E-4 & 5.47060E-4  \\\hline 
128& 2.52982E-5 & 1.19786E-5 & 1.04332E-5 & 3.65026E-5 & 1.35787E-4  \\\hline 
256& 6.29958E-6 & 2.98229E-6 & 2.59901E-6 & 9.09201E-6 & 3.38199E-5  \\\hline
\end{tabular}
\caption{\label{Tab:2d orth square MMSdataTable} $L_2$-norm error between
  computational and analytic MMS potentials for different grid
  resolutions and grid nonuniformities ($\epsgrid$) for
  the 2D orthogonal grid given by \eqn{periodic sin grid}. 
  For these tests, we have chosen $N_\xi = N_\eta$.}
\end{table}

For a unit square physical domain, we have chosen an MMS potential
given by
\begin{equation}
\Phi_\mathrm{MMS} = \sin(2 \pi x) \sin(2 \pi y),
\end{equation}
such that, assuming Cartesian geometry ($f=1$), the MMS charge density is
\begin{equation}
\rho^x_\mathrm{MMS} = 2 \pi \sin(2 \pi x) \sin(2\pi y).
\end{equation}
For this particular choice of $\Phi_\mathrm{MMS}$, we test our Poisson
solver with periodic field boundary conditions on all boundary segments.
Table~\ref{Tab:2d orth square MMSdataTable} shows the $L_2$-norm of
the error in ($\Phi - \Phi_\mathrm{MMS}$) as given by \eqn{L_2 Norm}
for the orthogonal, nonuniform grid (\eqn{periodic sin grid}) for
various levels of nonuniformity, $\epsilon_{g}$.  The
results scale with second-order accuracy in grid spacing as expected, even for the
$\epsgrid = 0.15$ case, in which the ratio of the area of the largest
to the smallest cells is $J_{\mathrm{max}}/J_{\mathrm{min}} =
\left(\frac{1+2\pi\epsgrid}{1-2\pi\epsgrid}\right)^2 \approx 1140$!

%\subsubsubsection{Nonorthogonal Grid}
\begin{figure}[h!]
\centering
\epsfig{file = 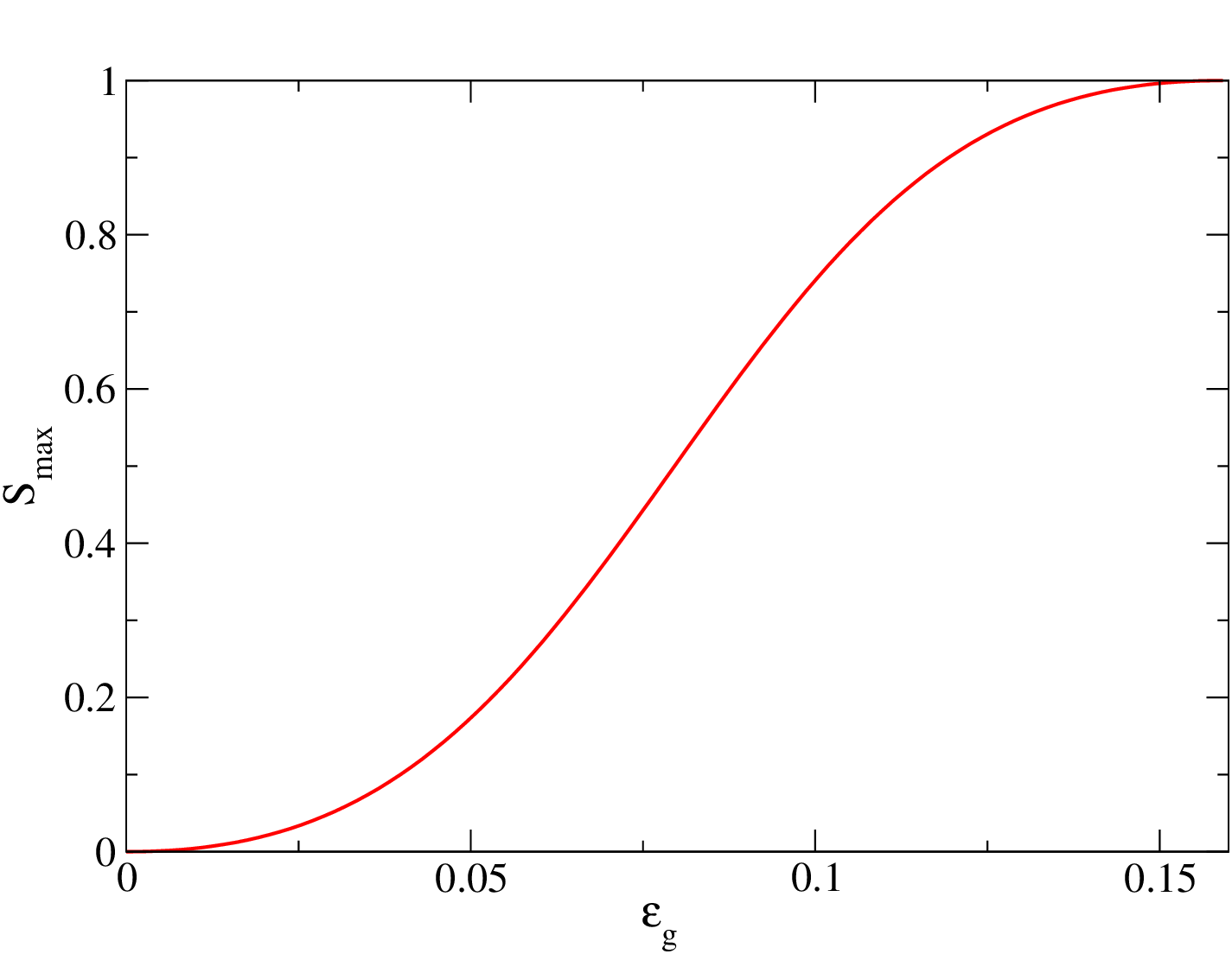,height=2.5in}
\caption{Graph of the maximum skewness parameter $S$ as defined in
\eqn{Eqn:skewness} as a function of $\epsgrid$ for the grid given by
\eqn{nonorthogonal grid}.}
\label{Fig:SkewnessParameter}
\end{figure}

\begin{table}[h]
\centering
\small
\centering
\begin{tabular}{|c|c|c|c|c|}\hline
  $\mm{N_\xi}$ & $\mm{ \epsilon_{g} = 0.025}$ & $
  \mm{\epsilon_{g} = 0.05}$ & $\mm{\epsilon_{g} =
    0.075}$& $ \mm{\epsilon_{g} = 0.15}$\\\hline\hline 
  16 & 2.08019E-3 &  3.27156E-3 & 5.57519E-3 & 8.99093E-3\\\hline 
  32 & 5.07340E-4 &  8.20596E-4 & 1.44430E-3 & 2.34622E-3\\\hline 
  64 & 1.25103E-4 &  2.03916E-4 & 3.62134E-4 & 5.90944E-4\\\hline 
  128& 3.10476E-5 &  5.07074E-5 & 9.02599E-5 & 1.47505E-4\\\hline 
  256& 7.73260E-6 &  1.26353E-5 & 2.25042E-5 & 3.67914E-5\\\hline
\end{tabular}
\caption{\label{Tab:2d nonorth square MMSdataTable} $L_2$-norm error between
  computational and analytic MMS potentials for different grid
  resolutions and grid nonuniformities ($\epsgrid$) for
  the 2D nonorthogonal grid given by \eqn{nonorthogonal grid}. For these tests, 
  we have again chosen $N_\xi = N_\eta$.}
\end{table}
Turning our attention to the nonorthogonal grid \eqn{nonorthogonal
  grid}, we must now also consider the amount of ``skewness'' of the
grid as a major factor in the difficulty associated with solving the
curvilinear Poisson equation \eqn{2d Logical Grid Poisson}. 
In an
effort to characterize the amount of nonorthogonality, or
``skewness,'' of our grids, we can use the Cauchy-Schwartz
inequality~\cite{BuckBook}, which states that any two points $\vv{p}$
and $\vv{q}$ in $n$-space must satisfy $\vv{p} \cdot \vv{q} \leq |\vv{p}||\vv{q}|$.
Writing the contravariant metric tensor in the form $g^{\alpha
  \beta}(\vv{x}\,) = \nabla{\xi}^\alpha \cdot \nabla{\xi}^\beta$,
the inequality can then be rewritten as
\begin{equation}\label{grid Schwartz Inequality}
\nabla \xi \cdot \nabla \eta \leq |\nabla \xi||\nabla \eta|.
\end{equation}
Squaring both sides and rearranging allows us to define the local grid
skewness factor
\begin{equation}\label{Eqn:skewness}
  S(\xi,\eta) = \frac{(\nabla \xi \cdot \nabla \eta)^2}
  {\left| \nabla \xi\right|^2 \left|\nabla \eta \right|^2} = 
  \frac{(g^{12})^2}{g^{11}g^{22}},
\end{equation}
where $0 \leq S \leq 1$.  In the limit of $S(\xi,\eta) \rightarrow 0$,
the grid is orthogonal at $(\xi,\eta)$, whereas for
$S(\xi,\eta)\rightarrow 1$, the grid becomes singular, i.e. the grid
cells become elongated and can fold.  Furthermore, we can define the
maximum grid skewness,
\begin{equation}\label{max skewness}
S_\mathrm{max} = \max\limits_{i,j} {\left[S(\xi_i,\eta_j)\right]},
\end{equation}
as a single parameter to characterize the maximum
nonorthogonality of the generated grid.
Figure~\ref{Fig:SkewnessParameter} shows $S_\mathrm{max}$ as a function
of $\epsgrid$ for the grid given by \eqn{nonorthogonal grid}.  For
$\epsgrid \geq 0.125$, $S_\mathrm{max}$ is very nearly unity, and thus
the grid is almost singular at some point on the grid.  As noted
above, this grid folds at $\epsgrid = \frac{1}{2\pi}\approx 0.16$.
As $S_\mathrm{max}\rightarrow 1$, the stiffness of the
Laplacian matrix becomes quite large and the field solver will not
easily converge, if at all.

The MMS tests performed on the nonorthogonal grid used the same MMS
potential as was used for the orthogonal square grid (and again $f=1$).
Table~\ref{Tab:2d nonorth square MMSdataTable} shows that the
$L_2$-norm of the error scales with second-order accuracy in $\Delta
\xi=\Delta\eta$, as expected.  We note here that the ratio of maximum cell area to
minimum for the grid defined in \eqn{nonorthogonal grid} scales
according to $\frac{J_{max}}{J_{min}} =
\frac{1+2\pi\epsgrid}{1-2\pi\epsgrid}$ ($\sim 34$ for $\epsgrid =
0.15$), meaning that the ratio of the largest cell area to the
smallest is much smaller than that of the orthogonal, nonuniform grid
case, but the skewness of the grid adds to the challenge and leads to
errors comparable to those in Table 1.

We have also designed an MMS test to validate our Poisson solver for
the annular grid generated in Section~\ref{sec:grid generation}.  We
simulate half the annulus and apply symmetry conditions at the bottom.
We apply either Dirichlet or Neumann boundary conditions along the
inner and outer boundary segments, and symmetry requires homogenous
Neumann boundary conditions along $\theta = 0,\pi$. If the boundary conditions consist of only
Neumann and periodic segments and no Dirichlet segments, the range of
the Laplacian operator consists of densities that have total charge
exactly equal to zero. Furthermore, this corresponds to a null space in
the Laplacian operator $\Phi \rightarrow \Phi + \mathrm{const}.$ We have found
that if we assure that the total charge is equal to zero, then the conjugate gradient
algorithm converges quickly.  The additional constant
potential does not affect the electric field. We have therefore
constructed the following potential:
\begin{equation}\label{phi_MMS}
\Phi\left(r,\theta \right) = 1 - r^3 + \left(r - r_1\right)\left(r_2 -
r\right)\cos\theta,
\end{equation}
where we have used polar coordinates to express \eqn{phi_MMS} in the
physical space, as it more naturally aligns with the annular grid case
than the Cartesian notation we have been using until this
point. At this point we should mention 

As our Poisson solver is designed to solve general geometries in the
physical space, we have applied the MMS test to both the azimuthally
symmetric and axisymmetric cases ($f=1$ and $f=r$, respectively.) The
physical Laplacian in this geometry with $f=1$, a cylindrical annulus,
is given by \eqn{Poisson azi}, leading to the physical charge density
\begin{equation}\label{rhortheta}
\rho^x\left(r,\theta\right) = \frac{1}{\fourpi} \left[9\,r + \left(3 -
\frac{r_1r_2}{r^2}\right)\cos\theta\right].
\end{equation}
This density is then multiplied by the Jacobian and inserted into the Poisson
solver with the proper boundary conditions.

\begin{table}[t!]
\centering
\subtable[Cylindrical Laplacian]{
\small
\centering
\begin{tabular}{|c|c|c|c|}\hline
$\mm{N_{\xi}}$ & $\mm{R = 5}$ & $\mm{R = 10}$ & $\mm{R = 20}$ \\\hline\hline
16  & 4.86684E-3 & 8.78720E-3 & 1.35298E-2 \\\hline
32  & 1.18612E-3 & 2.15682E-3 & 3.35229E-3 \\\hline
64  & 2.92357E-4 & 5.32567E-4 & 8.29718E-4 \\\hline
128 & 7.25471E-5 & 1.32213E-4 & 2.06105E-4 \\\hline
256 & 1.80677E-5 & 3.29310E-5 & 5.13433E-5 \\\hline 
\end{tabular}
\label{Tab:CylMMSdata}
}
\subtable[Spherical Laplacian]{
\small
\centering
\begin{tabular}{|c|c|c|c|}\hline
$\mm{N_{\xi}}$ & $\mm{R = 5}$ & $\mm{R = 10}$ & $\mm{R = 20}$ \\\hline\hline
16  & 3.84324E-3 & 6.41377E-3 & 9.23903E-003 \\\hline
32  & 9.43887E-4 & 1.60080E-3 & 2.35531E-003 \\\hline
64  & 2.33099E-4 & 3.96923E-4 & 5.87121E-004 \\\hline
128 & 5.78702E-5 & 9.86410E-5 & 1.46102E-004 \\\hline
256 & 1.44142E-5 & 2.45755E-5 & 3.64121E-005 \\\hline
\end{tabular}
\label{Tab:SphMMSdata}
}
\caption{$L_2$-norm error between computational and analytic MMS
  potentials for different grid resolutions and ratios $R \equiv
  r_2/r_1$ for the annular grids generated in 
  Section~\ref{sec:grid generation} for
  both Cartesian (a) and cylindrical (b) solvers. For these tests, we
  have again chosen $N_{\xi}= N_{\eta}$}
\label{Tab:MMSdataTable}
\end{table}

In the axisymmetric system ($f=r$), the spherical radius
satisfies $r_s^2 = r^2 + z^2$.  The physical domain consists of an
outer sphere of radius $r_{s2}$ is created in which an inner sphere of
radius $r_{s1}$ has been removed.  The corresponding physical charge
density is obtained via \eqn{Poisson axi},
\begin{equation}\label{rhosph}
\rho^x\left(r_s,\theta\right) = \frac{1}{\fourpi}\left[12\,r_s + \left(4 - \frac{2
r_{s1} r_{s2}}{r_s^2}\right)\cos\theta\right].
\end{equation}

\begin{figure}[t!]
\centering
\subfigure[\label{cylpcolor}]{
\includegraphics[width=2.5in]{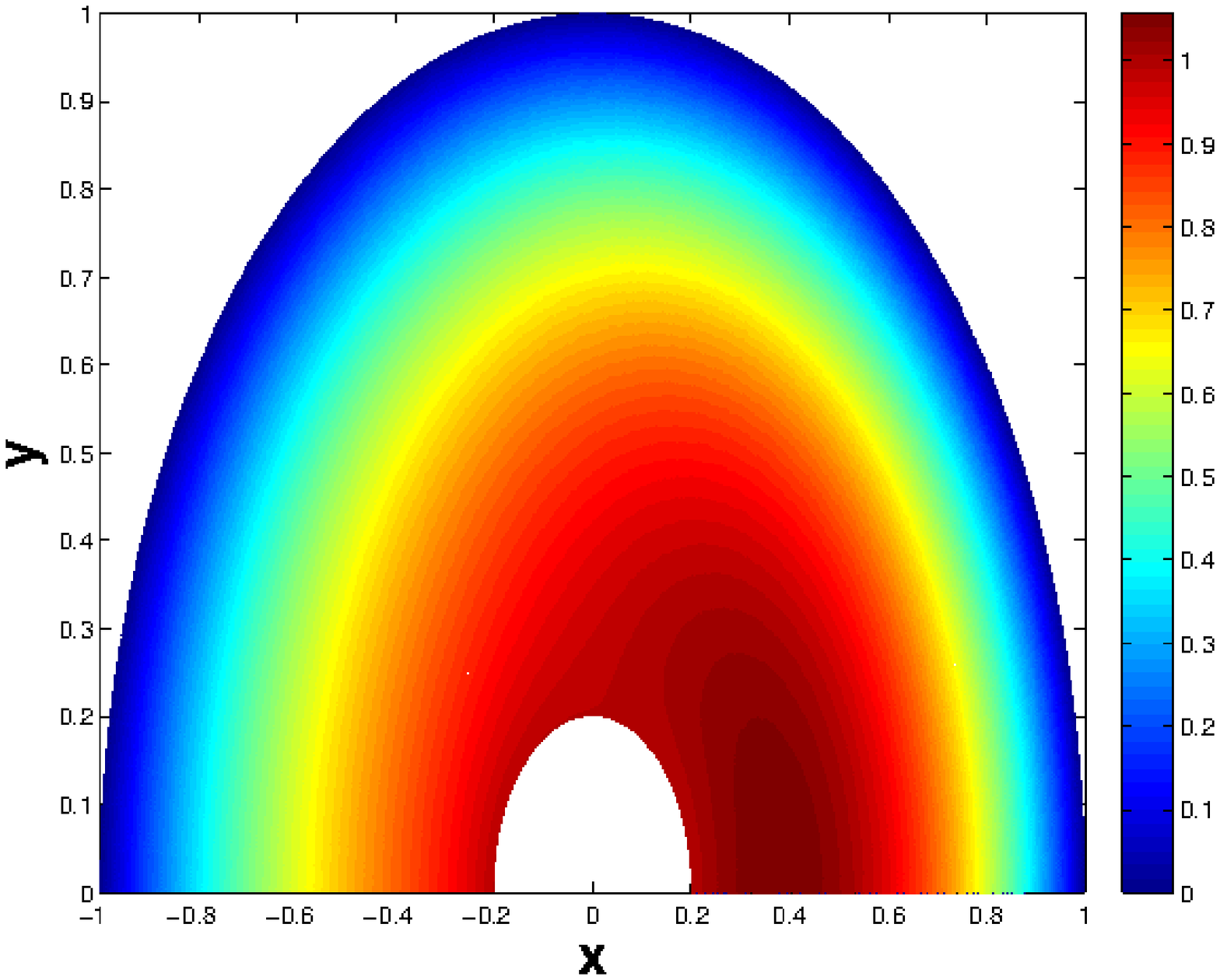}}
\subfigure[\label{cylmesh}]{
\includegraphics[width=2.5in]{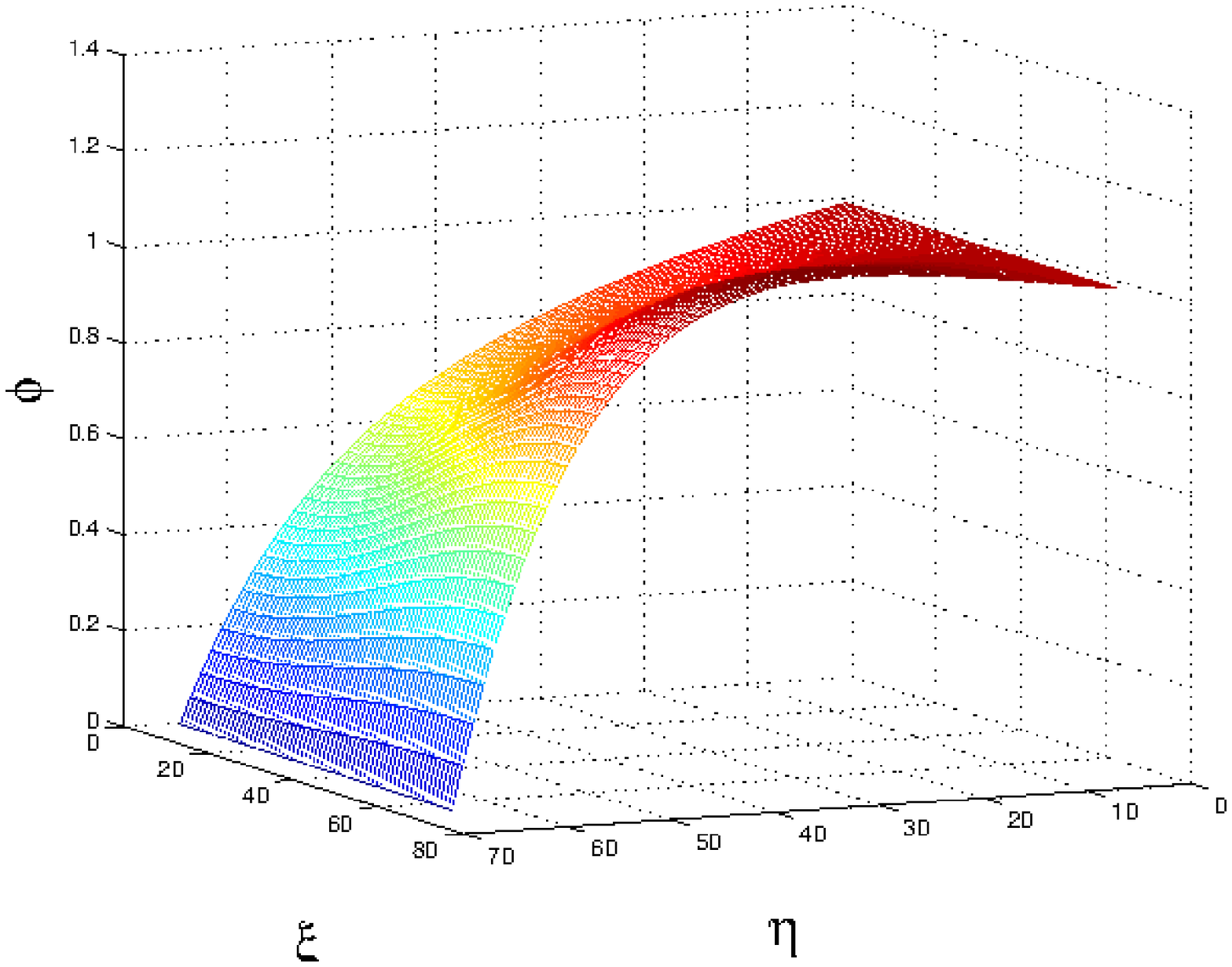}}
\caption{Two-dimensional MMS potential plots as obtained
computationally using (\ref{rhortheta}) and (\ref{rhosph}) in
physical (a) and logical (b) space.}
\label{Fig:MMSpotpcolor}
\end{figure}

The $L_2$-norm of the error for both azimuthal and axisymmetric
solvers is shown in Table~\ref{Tab:MMSdataTable}.  Here $R = r_2/r_1$
$\left(\mathrm{or}\, r_{s2}/r_{s1}\right)$ is the ratio of the radius
of the outer boundary to that of the inner boundary and $N_{\xi,\eta}$
is the number of uniformly spaced grid points in each direction.
Figure~\ref{Fig:MMSpotpcolor} displays the chosen MMS potential on
both the physical and logical domains as obtained by the logical grid
Poisson solver.  Notice that the different values of $\rho^\xi$
inserted in the solver for the different coordinate systems should
(and do) return the same potential on the grid.  The results shown in
Tables~\ref{Tab:2d nonorth square MMSdataTable}
and~\ref{Tab:MMSdataTable} show the expected second-order scaling with
grid spacing, showing that our field solver is working correctly.

\subsection{Self-Fields and Momentum Conservation}

With a uniform grid, symmetrical particle shapes guarantee exact
momentum conservation if the gridding is done
correctly~\cite{BirdsallLangdonBook}.  For a uniform rectangular grid,
this is equivalent to having the self-electric forces exactly zero
on each particle.  (\ref{Ham 2d mom update}) dictates that we use the
logical grid electric field at the particle's position on the logical
grid. The most obvious method to do this is by simply differencing
$\Phi$ with respect to $\xi$, (where we have defined a staggered mesh
such that $E^\xi$ exists on vertices and $\Phi$ and $\rho^\xi$ exist
on cell centers).  One then interpolates $E^\xi$ to the logical space
particle position.  Unfortunately, simple 1d tests on this method of
obtaining the logical electric field with a single particle reveal a
non-zero self-field at the particle.

We have therefore devised a second, less direct method of
obtaining the logical electric fields at the particle position in
which we solve for the electric field on the logical grid and
interpolate this field to the particle position. As such, we calculate
the electric field
on the vertices, formulated on the logical grid as
\begin{subequations} \label{2d E log from phi}
\begin{equation}\label{2d Exi}
  E^\xi_{i,j} =-\frac{\Phi_{i+\half,j+\half} - \Phi_{i-\half,j+\half} + 
\Phi_{i+\half,j-\half} -\Phi_{i-\half,j-\half}}{2 \Delta \xi}
\end{equation}
and
\begin{equation}\label{2d Eeta}
  E^\eta_{i,j} = -\frac{\Phi_{i+\half,j+\half} - \Phi_{i+\half,j-\half} + 
\Phi_{i-\half,j+\half} - \Phi_{i-\half,j-\half}}{2\Delta \eta}.
\end{equation}
\end{subequations}
We then apply the proper extrapolation techniques such that the
overall second-order accuracy of the system is upheld and calculate
the physical electric fields on the vertices using
\begin{subequations} \label{2d E phys}
\begin{equation}\label{2d Ex}
  E^x_{i,j} =\frac{1}{J^v_{i,j}}\left(j^v_{22,i,j} E^\xi_{i,j} -
  j^v_{21,i,j} E^\eta_{i,j} \right)
\end{equation}
and
\begin{equation}\label{2d Ey}
  E^y_{i,j} =\frac{1}{J^v_{i,j}}\left(j^v_{11,i,j} E^\eta_{i,j} -
  j^v_{12,i,j} E^\xi_{i,j} \right),
\end{equation}
\end{subequations}
where we have transformed from the inverse Jacobian matrix, $k^{\mu
  \nu}$ to the Jacobian matrix $j_{\mu \nu}$.  Here, the superscript
$v$ on the components of the Jacobi matrix and its determinant signify
that they are calculated on the vertices using a four-point average
from their natural cell-centered locations.  This technique for
calculating the electric fields at the particles leads to exactly zero
self-forces on a test particle in 1d, but in 2d the Jacobi matrix
components do not exactly cancel, leading to non-zero self-forces on
the particles.  Reference~\cite{dissertation} provides a detailed
explanation as to why this method leads to non-zero self-forces at the
particle and a detailed comparison of this method with the direct
interpolation method detailed above.  We stress that, with a non-uniform
grid, zero self-forces do not guarantee exact momentum conservation
because of the presence of the inertial terms in (\ref{Hamiltonian
  EOM}), which are often larger than the self-force
terms~\cite{dissertation}.

\subsubsection{Logical Electric Fields at a Particle}\label{2d E at part}

To test the self-forces on the particle in 2d, we have set up a system
in which a single particle is at rest on a doubly-periodic grid with a
neutralizing background such that we can interpolate both of the
physical electric fields and required grid derivatives to the grid
positions and retroactively multiply them to obtain the logical fields
at the particle position.  As mentioned above, we expect to have some
small self-force in 2d by utilizing this method.  Below we attempt to
quantify these forces.

In 2d, the logical electric fields are obtained from the physical
fields using
\begin{equation}\label{2d E log from E phys}
  \frac{\partial \Phi}{\partial \xi^\mu}= \frac{\partial x^\nu}
{\partial \xi^\mu}
  \frac{\partial \Phi}{\partial x^\nu}
\end{equation}
such that
\begin{equation}\label{2d E log from E phys}
\begin{array}{rcl}
E^\xi  &=& j_{11}\,E^x + j_{21}\,E^y \\
E^\eta &=& j_{12}\,E^x + j_{22}\,E^y.
\end{array}
\end{equation}
For a uniform grid, the electric fields at the particle are zero to
machine precision.  However, as soon as we allow one of the grid
dimensions to become nonuniform, the fields at the particle in the
nonuniform dimension are no longer zero, even for $\epsgrid =
10^{-4}$.  We find that the fields at the particle position scale with
second-order accuracy in grid spacing, and the magnitude of the field at
the particle in the nonuniform dimension is dependent upon the
magnitude of the grid nonuniformity parameter, $\epsgrid$.  For
example, using a grid with $N_\xi = N_\eta = 32$ where $\epsgrid =
0.15$ in $x$ and holding $y$ to be uniform ($\epsgrid = 0$),
$E^\xi(\xi_p,\eta_p) \approx 10^{-5}$, whereas $E^\eta(\xi_p,\eta_p)$
is zero to machine precision.  For $\epsgrid = 0.1$,
$E^\xi(\xi_p,\eta_p) \approx 10^{-7}$ for the same grid resolution.

We have also tried the direct interpolation method, i.e. directly
interpolating the logical electric field $E^{\xi}$ on the logical grid
to the particle position. This method also scales with second-order
accuracy in grid spacing, but the indirect method detailed above
consistently provides smaller fields at the particle position.

%%%%%%%%%%%%%%%%%%%%%%%%%%%%%%%%%%%%%%%%%%%%%%%%%%%%%%%%%%%%%%%%%%%%%%
%%%%%%%%%%%%%%%%%%%%%%%%%%%%%%%%%%%%%%%%%%%%%%%%%%%%%%%%%%%%%%%%%%%%%%

\section{NUMERICAL RESULTS} \label{sec:results} 

We have performed several benchmarking tests to validate the full 2D
curvilinear coordinate PIC code.  For all tests presented
below, we use a nonuniformity parameter of $\epsgrid = 0.1$ (for both
orthogonal and nonorthogonal grids). This leads to a ratio of areas
of the largest cell to smallest of $\sim 20$ for the nonuniform
orthogonal grid \eqn{periodic sin grid} and $\sim 4.4$ for the
nonorthogonal grid \eqn{nonorthogonal grid}.  The maximum skewness
parameter $S_\mathrm{max}$ is $\sim 0.75$ for the nonorthogonal
grid with this nonuniformity parameter.  We have done only high
frequency tests and have accordingly assumed that the ions are an
immobile background, providing charge neutrality in the
equilibrium. We have normalized the equations so that the electron
plasma frequency $\omega_{pe}$ is equal to unity.

\subsection{Cold Plasma Oscillations on a Square Physical Domain}\label{S:2d EPO}
\begin{figure}[t!]
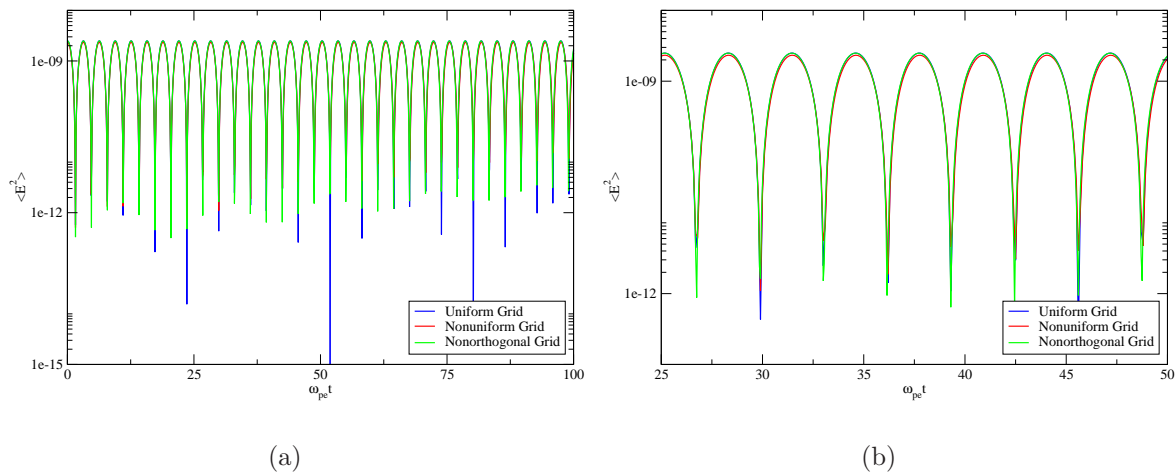

\centering \subfigure[]{
\includegraphics[width=3.in]{./figs/fig5a.eps}}
\subfigure[]{
\includegraphics[width=3.in]{./figs/fig5b.eps}}
\caption{Comparison
of cold plasma oscillation field energies using a uniform grid and
the nonuniform grids given by \twoeqns{periodic sin
  grid}{nonorthogonal grid} for (a) a long run and (b) a zoomed
section of the run showing an oscillation period of $2\pi$ for all
three grids.  Here we have used quadratic $(S_2)$ particle shape
functions with $N_\xi = N_\eta = 128$, $\bar{N}_\mathrm{ppc}=225$,
$\Delta t = 0.025$, and $\epsgrid = 0.1$ (for the nonuniform case).}
\label{Fig:2dESwave GR comp}
\end{figure}
Figure~\ref{Fig:2dESwave GR comp} shows a comparison of the evolution of
the electrostatic field energy, defined in 2d as
\begin{equation}\label{2d ES energy}
<E^2> = \frac{1}{2}\int dx\, dy \left[(E^x)^2 + (E^y)^2\right] ,
\end{equation}
for a cold plasma oscillation on the uniform, nonuniform but
orthogonal, and nonorthogonal grids.  For these tests, we take a cold
distribution of electrons with a uniform density stationary ion
background and initialize a perturbation in the system by perturbing
the particle positions with respect to the uniform neutralizing
background.  We define the quantity $\epsilon_\mathrm{pert}$ to be the
size of this initial perturbation on the particle positions.  The
initial particle positions, whether assigned uniformly in the physical
space or assigned according to the Jacobian in the logical space, lead
to a perturbed charge density. This perturbation can dominate
$\epsilon_\mathrm{pert}$ if the latter is small enough. For these
tests, the initial particle perturbation is directed at a $45^\circ$
angle across the grid to check the effects of the interpolation in
multiple dimensions on the data produced.  In Figure~\ref{Fig:2dESwave
  GR comp}, we show both a long-time evolution and a zoomed section of
the field energy from these same runs.  We have used a perturbation of
$\vv{\epsilon}_\mathrm{pert} = (7.07e^{-5},7.07e^{-5})$, such that the
magnitude of the perturbation is $|\vv{\epsilon}_\mathrm{pert}| =
1\times10^{-4}$, $N_\xi = N_\eta = 128$, the average number of
particles per cell is $\bar{N}_\mathrm{ppc}=225$, and
$\omega_{pe}\Delta t = 0.025$.  No measurable growth in the
electrostatic field energy was observed during the course of these
runs.  We note that the period of the plasma oscillations observed in
Figure~\ref{Fig:2dESwave GR comp} is very nearly $2 \pi$ (as is
expected for $\omega_{pe}=1$) for all three grid choices.  Further
testing has revealed that the period of the plasma oscillations
converges to the value of $2 \pi$ with second-order accuracy in
$\Delta t$, as expected, and the electric field energy does 
not begin to decay even in very long runs.

\subsection{Cold Electron-Electron Two-Stream Instability}\label{S: 2d 2stream}

\clearpage
\begin{figure}[h!]
\centering \epsfig{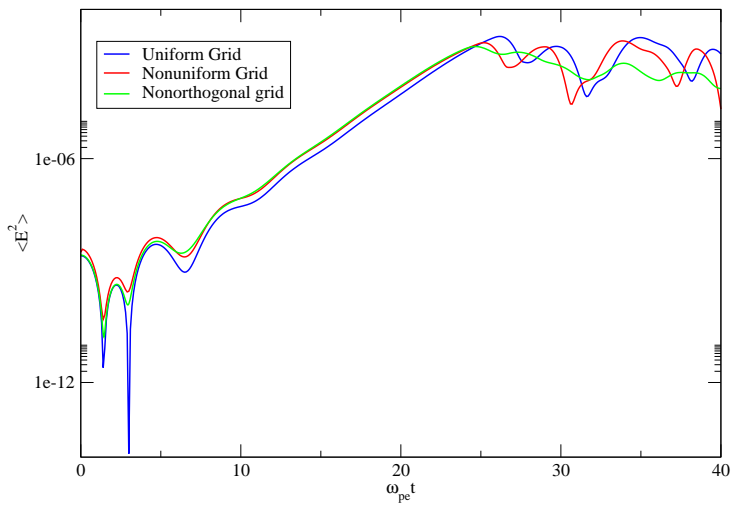}
\caption{Comparison of cold electron-electron two-stream instability
growth rates for uniform and non-uniform grids.  Higher initial
noise levels due in the nonuniform grid case provide a larger
initial perturbation in the system, leading to the differences seen
between the two curves.  Here we have used quadratic particle shape
functions with $N_\xi = N_\eta = 128$, $\bar{N}_\mathrm{ppc}=225$,
 $\Delta t = 0.025$ for all cases, with $\epsgrid = 0.1$
for the nonuniform orthogonal and nonorthogonal cases.}
\label{Fig:2d 2stream GR comp}
\end{figure}

Having chosen a physical grid such that $x,y \in [-\pi:\pi]$, the wave vector $\vv{k}$
is given by $\vv{k} = [1,1]$, such that, to generate a two stream
instability at a $45^\circ$ degree angle across the grid, we use the
effective wave vector, $k^{45^\circ} = \sqrt{2} \approx 1.414$.  In
our normalized units, the maximum growth rate for the two stream
instability of $\sim 0.35$ occurs at $|\vv{k}\cdot \vv{v}_0| \approx
0.63$.  As such, we have chosen an initial velocity parallel to
$\vv{k}$ (in the physical space) with $\vv{v} = [0.314,0.314]$.  

Figure~\ref{Fig:2d 2stream GR comp} shows the time evolution of the
electrostatic field energy for a uniform grid compared with the
nonuniform grids given by \twoeqns{periodic sin grid}{nonorthogonal
grid}.  The
offset of the three curves is due to the larger initial perturbation
due to the grid non-uniformity discussed above.
Here we have used quadratic particle shape functions with
$N_\xi = N_\eta = 128$, $\bar{N}_\mathrm{ppc}=225$,
$\omega_{pe}\Delta t = 0.025$ for all cases, again with $\epsgrid = 0.1$ for the
nonuniform orthogonal and nonorthogonal cases.  The same cases have
been performed using bilinear particle shape functions, revealing
identical results.  For all cases, the observed growth rate is $\sim 0.35$,
which matches the theoretical prediction.

\subsection{Landau Damping}\label{S:2d Landau Damping}

\afterpage{\clearpage
\begin{figure}[h]
\centering \epsfig{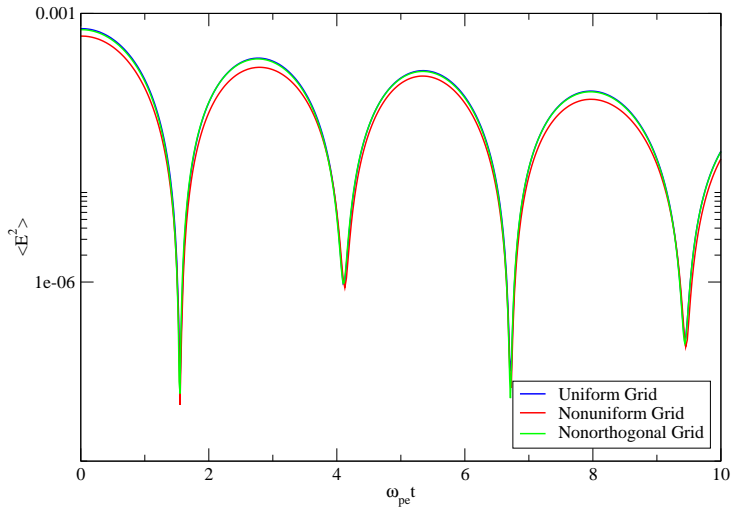}
\caption{Comparison of Landau damping rates for a uniform grid with
those obtained using the nonuniform, orthogonal and nonorthogonal
square grids given by \twoeqns{periodic sin grid}{nonorthogonal
  grid}.  Here we have used $N_\xi = N_\eta = 128$,
$\bar{N}_\mathrm{ppc}=400$ per species, $\Delta t = 0.025$, $v_{th}
= 0.07$, and quadratic particle shape functions for all cases, with
$\epsgrid = 0.1$ for the nonuniform orthogonal and nonorthogonal
cases.}
\label{Fig:2d LD GR comp}
\end{figure}}

In this section we study the effects of Landau damping with our 2D
code, on a unit square physical grid.  Since the our thermal
distribution is taken to be Maxwellian, the particle velocities are
isotropic in $x,y$, and as such we have set up a case in which the
initial perturbation is only in $x$ for simplicity.
Figure~\ref{Fig:2d LD GR comp} shows the time evolution of the
Landau damping on our electrostatic field energy for the uniform,
nonuniform orthogonal and nonorthogonal square grids.  Here we have
used $N_\xi = N_\eta = 128$, $\bar{N}_\mathrm{ppc}=400$, $\Delta t =
0.025$, and $v_{th} = 0.07$ for all cases.  Note that the
nonorthogonal grid more closely follows the damping rate of the
uniform grid than does the nonuniform, orthogonal grid.  The real
oscillation frequency agrees very well with the theoretical value of
$0.58$ given by the Bohm-Gross dispersion
relation~\cite{KrallandTrivelpieceBook} $\omega^2 = \omega_{pe}^2 + 3
k^2 v_{th}^2$, where again we have taken $\omega_{pe} = 1$ in
accordance with our code normalizations.  The average damping rate
across the simulation time agrees to within $3\%$ of the theoretical
value of $0.124$ as given by~\cite{KrallandTrivelpieceBook}
\begin{equation}\label{LandauDampingEqn}
  \omega_i =
  -\sqrt{\frac{\pi}{8}}\frac{\omega_{pe}}{|k^3\lambda_{De}^3|}
  \exp{\left[-\left(\frac{1}{2k^2
          \lambda_{De}^2}+\frac{3}{2}\right)\right]}
\end{equation}
for these parameters for the uniform and nonorthogonal grids, whereas
the damping rate on nonuniform, orthogonal grid agrees to within
$5\%$.  We hypothesize that the lower damping rate observed for the
nonuniform orthogonal grid case \eqn{nonorthogonal
  grid} is due to the strong nonuniformity of the
grid; the ratio of the largest to the smallest cell areas for a
nonuniformity parameter $\epsgrid = 0.1$ is $\sim 20$.  To check this
hypothesis, we have also run cases in which we have used $\epsgrid =
0.06$ such that the largest to smallest cell area ratio is $\sim 4.9$.
These cases give the same damping rates as the uniform grid case shown
above.

\subsection{Cold Plasma Oscillations on a Concentric Annulus}\label{S:2d EPO annulus}
\begin{figure}[tp!]
\centering
\epsfig{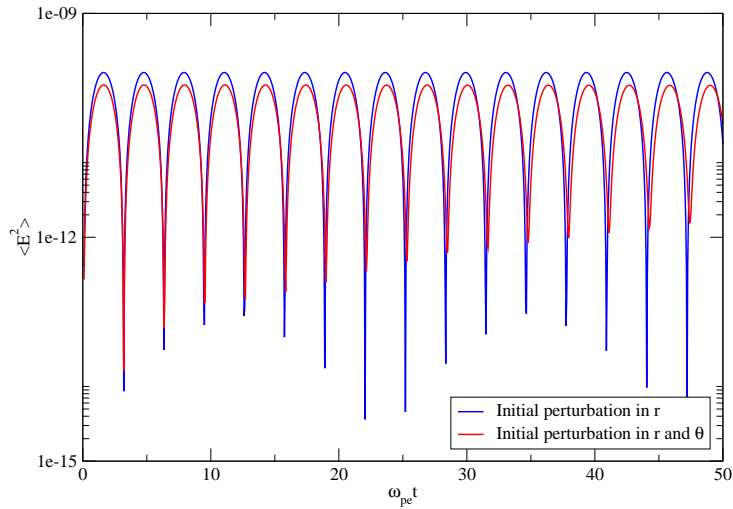}
\caption{Comparison of cold plasma oscillation field energies on an
  annular physical grid using an initial perturbation in $r$ only and
  a combination of $r$ and $\theta$ initial perturbation showing an
  oscillation period very close to $2\pi$ for both cases.  Here we
  have used quadratic spline particle shape functions with $N_\xi =
  N_\eta = 64$, $\bar{N}_\mathrm{ppc}=400$, and $\Delta t = 0.05$.}
\label{Fig:2dESwave GR comp Ann}
\end{figure}

As a final test of our entire method, we have set up a cold plasma
oscillation on the concentric annulus grid of Figure~\ref{Fig:Grids}a
(taking azimuthal symmetry).  As an initial test, we perturbed the
initial potential radially using
\begin{equation}\label{pert r only}
\tilde{\Phi} = \epsilon_\mathrm{pert}\cos{\left(\pi\left(\frac{r - r_1}{r_2 -
      r_1}\right)\right)},
\end{equation} 
where we have used $\epsilon_\mathrm{pert} = 10^{-4}$, $r_1 = 0.25$,
and $r_2 = 1.0$.  This initial perturbation satisfies homogeneous
Neumann boundary conditions along the entire boundary of the system.

In general, for Neumann boundary conditions, there is a subtlety,
namely that the unit normal to the physical dommain does not map to
the unit normal in the logical domain. Thus, Neumann boundary
conditions involve the normal and tangential derivatives on the
logical boundary. For this case, however, the Winslow coordinates are
orthogonal, and this issue does not arise. See Sec.~6.2 for a
discussion of the associated null space issues.

With our input parameters, the ratio of the area of the largest grid
cell to the smallest is $\sim 16.$ This is a very simple test of the
system, and is analogous to perturbing our system only in $y$ on a
rectangular grid.  The time evolution of the electrostatic field
energy for this test is shown in the blue curve of
Figure~\ref{Fig:2dESwave GR comp Ann} for a $64\times 64$ physical grid with
$\omega_{pe}\Delta t = 0.1$, $\bar{N}_\mathrm{ppc} = 225$, and quadratic spline
shape functions.  The period of oscillation of the field energy for
this set of initial conditions is observed to be $\sim 6.3$ for this case,
meaning that the frequency is indeed unity with an error that scales
as $\Delta \xi^2$, as per our code normalizations.

As a more challenging case, we then perturbed the initial potential
using
\begin{equation}\label{pert r theta}
\tilde{\Phi} =
  \epsilon_\mathrm{pert}\cos{\left(\pi\left(\frac{r - r_1}{r_2 -
          r_1}\right)\right)}\cos\theta, 
\end{equation}
where we have again used $\epsilon_\mathrm{pert} = 10^{-4}$, $r_1 =
0.25$, and $r_2 = 1.0$.  The red curve in Figure~\ref{Fig:2dESwave GR
  comp} shows the time evolution of the electrostatic field energy for
a $64\times 64$ physical grid with $\omega_{pe}\Delta t = 0.05$,
$\bar{N}_\mathrm{ppc} = 400$, and quadratic spline shape functions.
We have used more resolution in this particular case in order to
resolve more fully the complicated features of our initial potential.
Again, the measured period of oscillation is $\sim 6.3$ (the difference
between the two curves is approximately $0.085\%$); giving frequency
unity, with errors scaling as $\Delta \xi^2$.

 \begin{figure}[hbt!]
   \centering
   \subfigure[$\omega_{pe}\,t = 1.6$]{
     \includegraphics[width=2.85in]{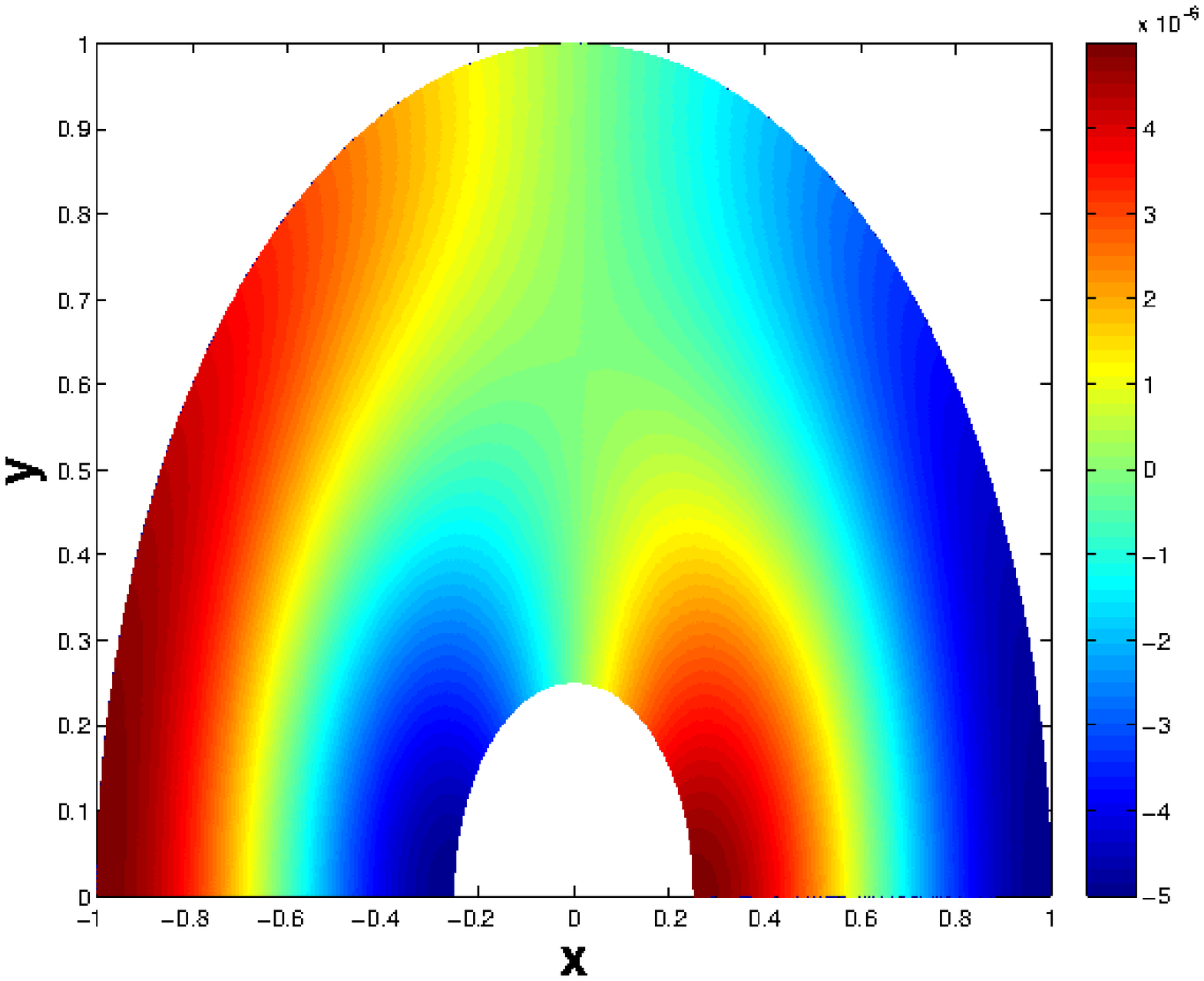}}
   \subfigure[$\omega_{pe}\,t = 2.8$]{
     \includegraphics[width=2.85in]{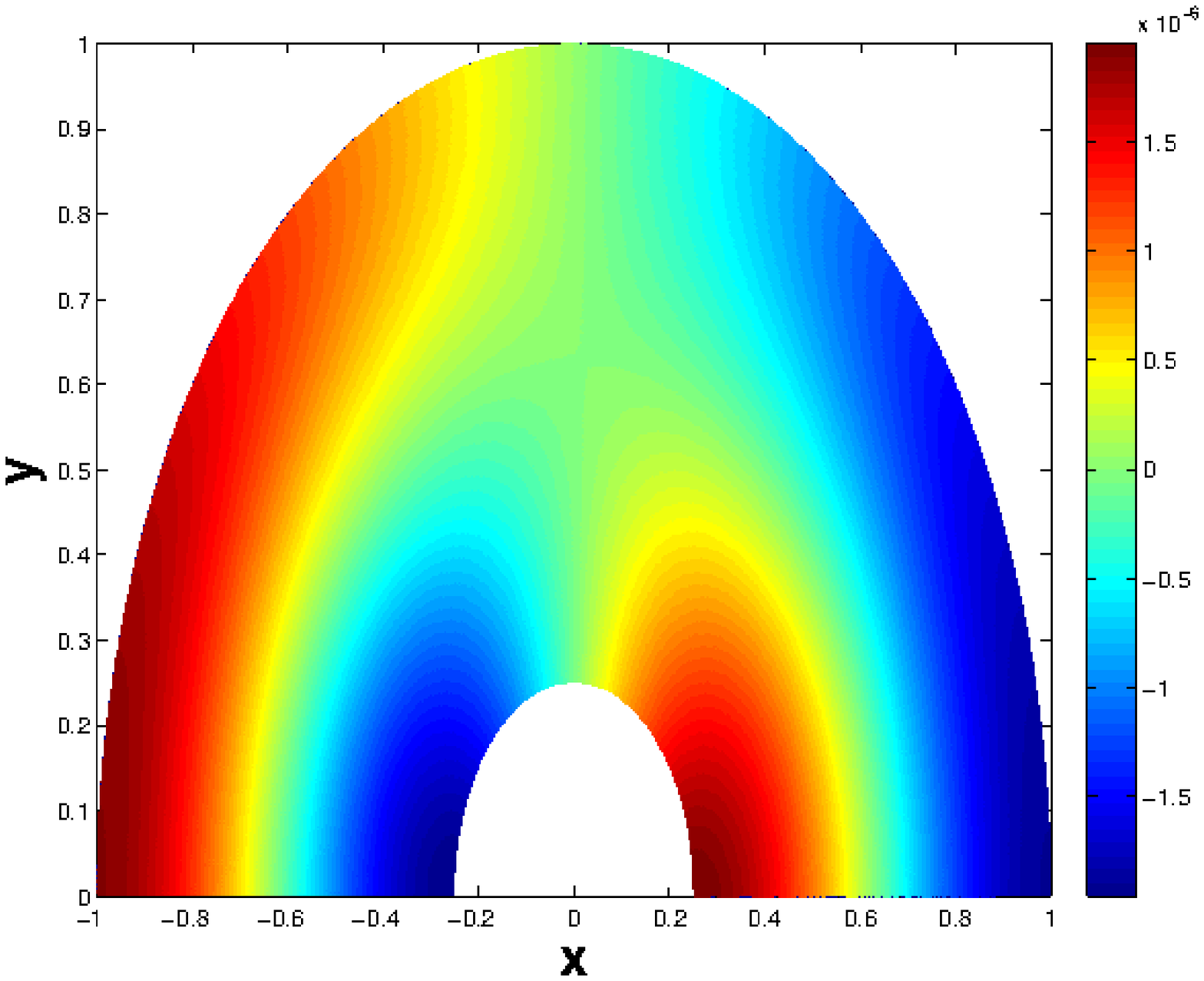}}
   \subfigure[$\omega_{pe}\,t = 3.6$]{
     \includegraphics[width=2.85in]{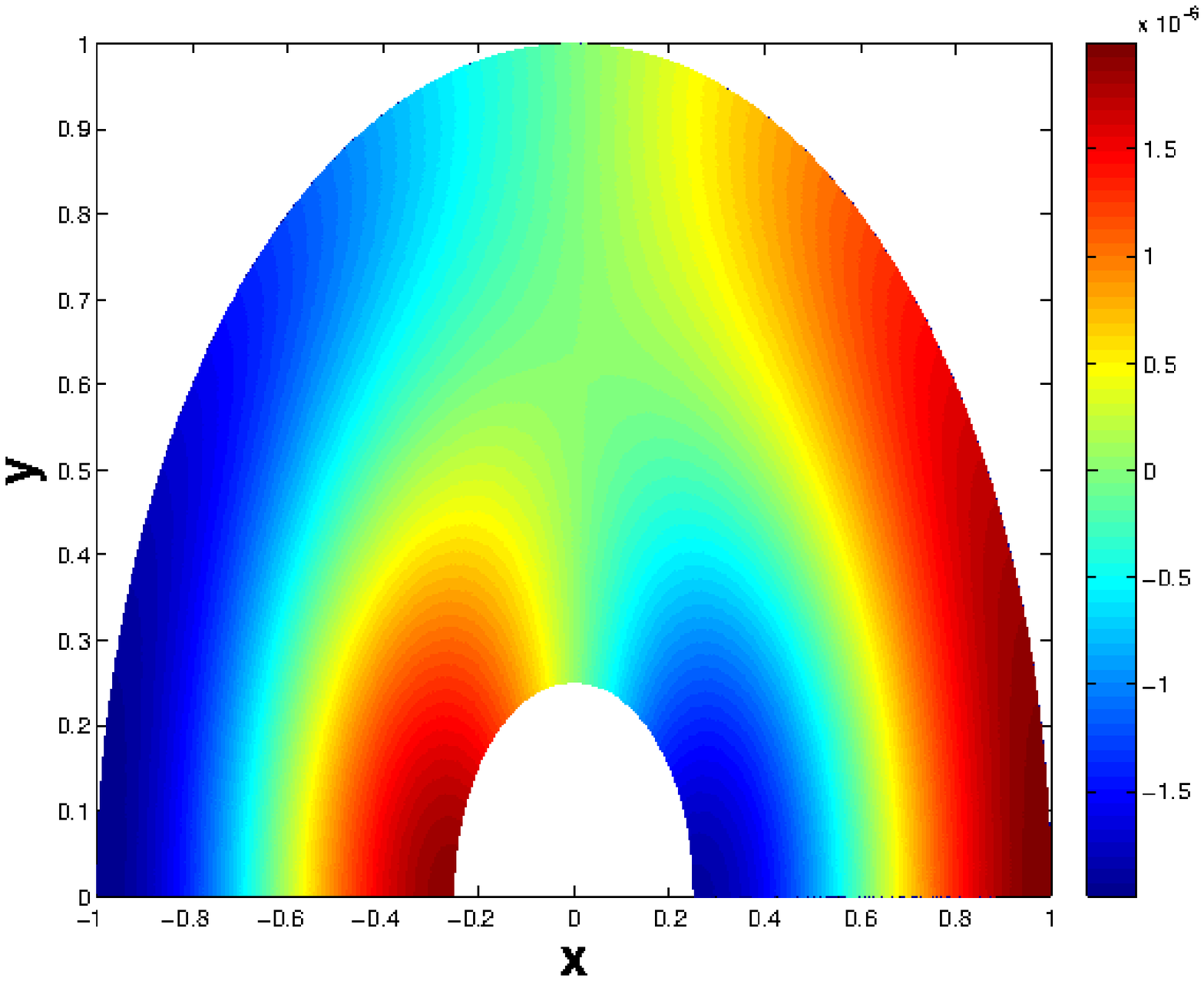}}
   \subfigure[$\omega_{pe}\,t = 4.8$]{
     \includegraphics[width=2.85in]{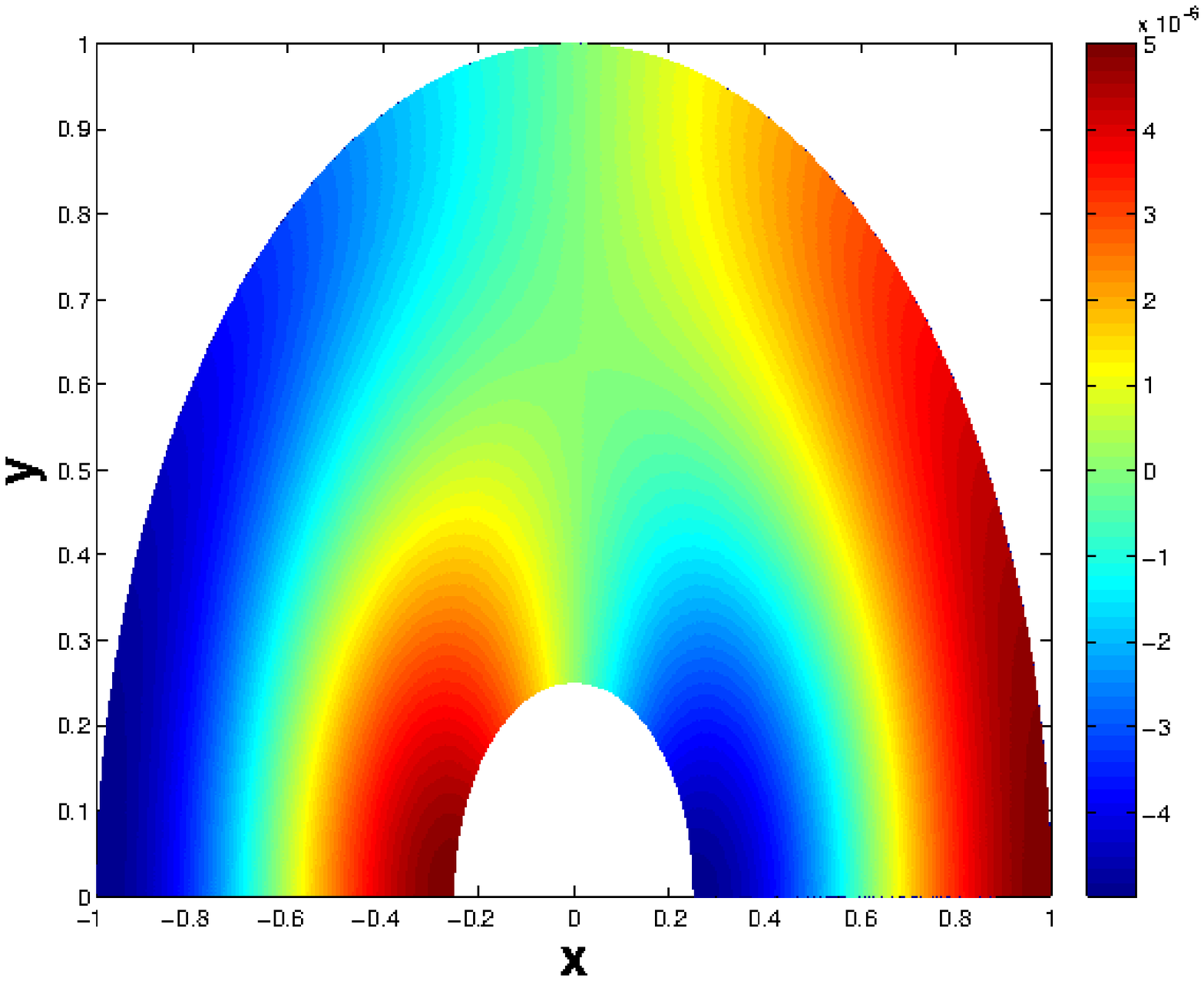}}
   \caption{Snapshots of one period of the evolution the potential of a
     cold plasma oscillation on the annular physical grid. Here we have
     perturbed the inital potential using \eqn{pert r theta}.}
   \label{Fig:Ann_ES_evolution}
 \end{figure}

 Figure~\ref{Fig:Ann_ES_evolution} shows the time evolution of one
 period of the potential on the physical grid using the initial
 perturbation as given by \eqn{pert r theta} for the cold
 electrostatic plasma oscillation test described above.
 Figures~\ref{Fig:Ann_ES_evolution}(a) and (b) are from one half of
 the plasma period and Figures~\ref{Fig:Ann_ES_evolution}(c) and (d)
 are from the next.  Notice the exact reversal of the potentials
 between the two halves of the plasma period.  Reflecting boundary
 conditions for particles at the non-periodic parts of the boundary
 were not required for this case, because the particle excursions for
 a cold plasma oscillation are so small.

%%%%%%%%%%%%%%%%%%%%%%%%%%%%%%%%%%%%%%%%%%%%%%%%%%%%%%%%%%%%%%%%%%%%%%
%%%%%%%%%%%%%%%%%%%%%%%%%%%%%%%%%%%%%%%%%%%%%%%%%%%%%%%%%%%%%%%%%%%%%%

\section{CONCLUSIONS} \label{sec:conclusions}
In this study, we have
set out to develop a new approach to the PIC method, in which the key
components, the mover, the field solve, charge
accumulation, and field interpolation, are carried out on a uniform
grid in the logical domain, the unit square, and mapped to an
arbitrary boundary conforming grid on an arbitrary physical domain. We
have focused on a 2d, electrostatic model as a proof of principle.

To demonstrate our method, we used analytically prescribed grids and 
grids obtained by Winslow's grid generation
technique~\cite{WinslowJCP} to generate a
boundary-fitted grid, in the latter case by solving a set of coupled elliptic equations
using a nonlinear Newton-Krylov solver. The generated grids are then
mapped onto the logical grid through the use of the mapping
$\vv{x}(\vv{\xi}\,)$ and its inverse, $\vv{\xi}(\vv{x})$, such that
the PIC components can be run on the logical grid using these
mappings.

We have derived the logical grid macroparticle equations of motion
based on a canonical transformation of Hamilton's equations from the
physical domain to the logical.  The resulting nonseparable system of
equations using an extended form of the semi-implicit modified
leapfrog integrator~\cite{FinnChacon05}, which we have shown to be symplectic for a system
of arbitrary dimension and second-order accurate in time if
symmetrization leading to a time-centered discretization of the
macroparticle update equations is utilized. If the field solve is performed
just after the first step of this integrator, it needs to be done only once per time step.

In order to obtain the electrostatic fields on the logical grid, we
have constructed a generalized curvilinear coordinate formulation of
Poisson's equation which is discretized conservatively on the logical
grid using a staggered mesh.  Field boundary conditions are applied in
such a way as to produce a symmetric operator matrix which we then
solve using a conjugate gradient solver.

Our formulation of the curvilinear coordinate Poisson equation
requires the logical grid charge density, which allows us to
accumulate the charge from the particles directly onto the uniform,
square logical grid using standard particle shape functions rather
than the more complicated, weighted shape functions which must be used
if the charge is accumulated on a nonuniform physical grid.
Furthermore, the particle equations of motion require that the
derivative of the electrostatic potential on the logical grid be
obtained at the particle positions for the update of the particle
momentum.  These logical electric fields are interpolated to the
particle positions on the logical grid using the symmetric particle
shape functions which have been slightly modified from the standard
shape functions used in the charge accumulation process in order to
account for our choice of a staggered mesh.  All validation tests
performed have shown that the code produces correctly the physics for
complicated interior meshing, as well as for complicated domain
shapes.

Albeit at a proof-of-principle level, our code has shown that an
arbitrary coordinate PIC method is a an accurate and efficient
alternative to current methods for simulating plasma systems with
complex domains.  Future work will focus on extending the method to 3d
and extending to electromagnetic systems.  Further, work is required
on parallelization techniques to handle more efficiently the particle
push and charge accumulation stages of our method.  Finally, our
method can be coupled to a moving mesh algorithm to more accurately
follow the dynamic evolution of the plasma system.

%%%%%%%%%%%%%%%%%%%%%%%%%%%%%%%%%%%%%%%%%%%%%%%%%%%%%%%%%%%%%%%%%%%%%%%
%%%%%%%%%%%%%%%%%%%%%%%%%%%%%%%%%%%%%%%%%%%%%%%%%%%%%%%%%%%%%%%%%%%%%%%
\ack The authors wish to thank L.~Chacon, V.~Roytershteyn and G.~L.~Delzanno for many helpful 
discussions.
The work of C.A.F. and J.M.F. was supported by the Laboratory Directed
Research and Development program (LDRD), under the auspices of the US
Department of Energy by Los Alamos National Laboratory, operated by
Los Alamos National Security LLC under contract
DE-AC52-06NA25396.   The work of K.C. was supported by the Air Force Research Laboratory, AFOSR and Sandia
National Laboratories (Sandia is a multiprogram laboratory operated
by Sandia Corporation, a Lockheed Martin Company, for the United States
Department of Energy's National Nuclear Security Administration under
Contract DE-AC04-94AL85000).
%%%%%%%%%%%%%%%%%%%%%%%%%%%%%%%%%%%%%%%%%%%%%%%%%%%%%%%%%%%%%%%%%%%%%%%
%%%%%%%%%%%%%%%%%%%%%%%%%%%%%%%%%%%%%%%%%%%%%%%%%%%%%%%%%%%%%%%%%%%%%%%

\appendix                                                 %%%% APPENDICES %%%%
\section{Differential Geometry Notation}\label{App:DifferentialGeometryNotation}
This Appendix presents the reader with a coherent overview of the
relations between Cartesian and curvilinear coordinates which form the
basis of this paper.

Let the values $x^\alpha , \alpha = 1,\cdots,n$ be the Cartesian
coordinates of the vector ${\vv x}$. The coordinate transformation
${\vv x}(\vv \xi\,)$ defines a set of curvilinear coordinates
$\xi^\alpha ,\cdots,\xi^n$ in the domain $X^n$.  We define the Jacobi
matrix of this transformation \begin{equation} \label{Jacobi matrix}
j_{\alpha \beta}(\vv{\xi}\,) \equiv \left(\frac{\ptl x^\alpha }{\ptl
    \xi^\beta}\right),\quad \alpha,\beta = 1,\cdots,n,
\end{equation}
and its Jacobian
\begin{equation} \label{jacobian}
J(\vv{\xi \,}) \equiv \mbox{det}\left(\mm j\right).
\end{equation}
Conversely, we can also think of this transformation as a mapping of
$\vv{\xi}$ to $\vv{x}$, $\vv \xi(\vv{x}\,)$. Defining the inverse of
the matrix $j_{\alpha \beta}$ as
\begin{equation} \label{inv Jacobi matrix}
k^{\alpha \beta} (\vv{x}\,)\equiv \left(\frac{\ptl
\xi^\alpha }{\ptl x^\beta}\right),\quad \alpha,\beta = 1,\cdots,n,
\end{equation}
we can write its Jacobian 
\begin{equation} \label{inv jacobian}
K(\vv{x}) \equiv \mbox{det}\left(\mm k\right) = \frac{1}{J}.
\end{equation} 
Simple linear algebra tells us that
\begin{equation} \label{delta fn}
j_{\alpha \beta} k^{\beta \gamma} = \frac{\ptl x^\alpha }{\ptl \xi^\beta}\frac{\ptl
\xi^\beta}{\ptl x^\gamma} \equiv \delta_\alpha^\gamma,
\end{equation}
therefore $\mm{j} \mm{k} = \mm{I}$.

Given a Euclidean metric on the physical space, we have
\begin{equation}
dx^\gamma dx^\gamma = \frac{\ptl x^\gamma}{\ptl \xi^\alpha }\frac{\ptl
x^\gamma}{\ptl \xi^\beta} d\xi^\alpha d\xi^\beta = g_{\alpha
\beta}d\xi^\alpha d\xi^\beta,
\end{equation}
and the covariant metric tensor is defined as
\begin{equation} \label{cov metric}
g_{\alpha \beta}(\vv{\xi}\,) \equiv \frac{\ptl x^\gamma}{\ptl \xi^\alpha }\frac{\ptl
x^\gamma}{\ptl \xi^\beta},\quad \alpha,\beta,\gamma = 1,\cdots,n.
\end{equation}
From (\ref{Jacobi matrix}), we see $g_{\alpha
\beta}=j_{\gamma\alpha} j_{\gamma\beta}$, which is simply $\mm
g_{\mathrm{cov}}=\mm j^T \mm j$.  Likewise, the contravariant metric
tensor is based upon the inverse Jacobian matrix, $\mm k$:
\begin{equation}\label{contra metric}
g^{\alpha \beta}(\vv{x}\,) \equiv \frac{\ptl \xi^\alpha }{\ptl x^\gamma}\frac{\ptl
\xi^\beta}{\ptl x^\gamma},\quad \alpha,\beta,\gamma = 1,\cdots,n,
\end{equation}
thus $\mm g^{\mathrm{con}}=\mm k \mm k^T$.  

Finally, by multiplying the covariant and contravariant metric
tensors and applying the identity $\mm j \mm k = \mm I$, we can prove
that the covariant and contravariant metric tensors are in fact inverses of
each other: 
\begin{equation} \label{cov contra mult}
\mm g_{\mathrm{cov}} \mm
g^{\mathrm{con}}=\mm j^T \mm j \mm k \mm k^T = \mm j^T \mm k^T =
\left(\mm k \mm j\right)^T = \mm I^T=\mm I.
\end{equation}
In two dimensions, we can easily convert from the covariant to the
contravariant metric tensor using the following equation:
\begin{equation}\label{App:cov to cont} 
g^{\alpha \beta}=\left(-1\right)^{\alpha+\beta}\frac{g_{3-\alpha,3-\beta}}{g_{\mathrm{cov}}},
\quad \alpha,\beta = 1,2,
\end{equation}
where $g_{\mathrm{cov}} = J^2$ is the determinant of the covariant
metric tensor, and $g_{\mathrm{cov}} = \frac{1}{g^{\mathrm{con}}}$,
where $g^{\mathrm{con}} = K^2$ is the determinant of the
contravariant metric tensor.  Likewise, we can shift from the
contravariant to the covariant metric tensor using
\begin{equation}\label{cont to cov} 
g_{\alpha \beta}=\left(-1\right)^{\alpha+\beta}g_{\mathrm{cov}}g^{3-\alpha,3-\beta},
\quad \alpha,\beta = 1,2.
\end{equation}
%%%%%%%%%%%%%%%%%%%%%%%%%%%%%%%%%%%%%%%%%%%%%%%%%%%%%%%%%%%%%%%%%%%%%%%
%%%%%%%%%%%%%%%%%%%%%%%%%%%%%%%%%%%%%%%%%%%%%%%%%%%%%%%%%%%%%%%%%%%%%%%
%
%\section{Derivation of Winslow's Method}\label{App: Winslow}
%
%%%%%%%%%%%%%%%%%%%%%%%%%%%%%%%%%%%%%%%%%%%%%%%%%%%%%%%%%%%%%%%%%%%%%%%
%%%%%%%%%%%%%%%%%%%%%%%%%%%%%%%%%%%%%%%%%%%%%%%%%%%%%%%%%%%%%%%%%%%%%%%

\section{Derivation of Curvilinear Coordinate Poisson Equation}\label{App: Poisson}
We begin by rewriting (\ref{2d Poisson Eqn}) as the divergence of
the gradient of the potential in generalized coordinates: 
\begin{equation} \label{poisson exp} 
\frac{1}{f}\nabla \cdot f\nabla \Phi = \frac{1}{f}\frac{\ptl }{\ptl
x^\alpha } \cdot f\frac{\ptl \Phi}{\ptl x^\alpha } =
-4\pi\rho,
\end{equation} 
 where $f$ is a geometry factor allowing us to switch between
azimuthal and axially-symmetric systems. For instance, if we want the
physical coordinate system to be $x,y$ with $\frac{\partial}{\partial
  z}=0$, we set $\mathit{f} = 1$. For $r,z$ coordinates with
$\frac{\partial}{\partial \phi}=0$, we set $\mathit{f} = r.$

In matrix notation we can write $\nabla \Phi$ as
\begin{eqnarray}\label{grad phi matrix} \nonumber 
\nabla \Phi &=& \frac{\ptl{\Phi}}{\ptl{\xi^m}} \nabla \xi^m \\ &=& A_m
\nabla \xi^m \end{eqnarray} where $A_m$ are the covariant components
of the 2D vector
\begin{equation} \label{A covariant} \vv{A}= A_m
\nabla \xi^m = A_1 \nabla \xi + A_2 \nabla \eta.  
\end{equation}
Likewise, we can represent $\vv{A}$ by its contravariant components,
$A^m$: 
\begin{equation} \label{A contravariant} \vv{A}= A^1 \nabla \eta
\times \hat{z} + A^2 \hat{z} \times \nabla \xi,
\end{equation}
such that we can write 
\begin{equation} \label{A_matrix} A_1 \nabla
\xi + A_2 \nabla \eta= A^1\nabla \eta \times \hat{z} + A^2 \hat{z}
\times \nabla \xi.
\end{equation} 
We can now find the direct relationship between the covariant and
contravariant components of $\vv{A}$.  For example, taking the inner
product of \eqn{A_matrix} with $\nabla \eta \times \hat{z}$ we find:
\begin{equation} \label{log poisson cov cont relation} A^1
\left(\nabla \eta \times \hat{z} \right) \cdot \nabla \xi = A_1
|\nabla \xi |^2 + A_2 \nabla \eta \cdot \nabla \xi.  
\end{equation} 
By (\ref{contra metric}), 
\begin{equation}\nabla \xi^\beta \cdot \nabla
\xi^\gamma = g^{\beta \gamma}, 
\end{equation}
allowing us to rewrite (\ref{log poisson cov cont relation}) as
\begin{equation}\label{A cov to cont conv}
A_1 g^{11} + A_2 g^{12} = A^1 \left(\nabla \eta \times
\hat{z} \right) \cdot \nabla \xi = A^1 \nabla \xi \cdot \nabla \eta
\times \hat{z}. 
\end{equation} 
We note that 
\begin{eqnarray}\label{J from cross product}\nonumber
\nabla \xi \cdot \nabla \eta \times \hat{z} &=& \frac{\ptl \xi}{\ptl
x}\frac{\ptl \eta}{\ptl y} - \frac{\ptl \xi}{\ptl y}\frac{\ptl
\eta}{\ptl x} \\ \nonumber &=& \det{(\mm{k})} \\ &=& \frac{1}{J},
\end{eqnarray}
so we can write the left-hand side of (\ref{A cov to cont conv}) as
\begin{subequations}
\begin{equation}
A_1 g^{11} + A_2 g^{12} = \frac{A^1}{J}. 
\end{equation} 
Similarly,
\begin{equation}
A_1 g^{21} + A_2 g^{22} = \frac{A^2}{J}. 
\end{equation} 
\end{subequations}

Returning now to (\ref{poisson exp}) and using (\ref{grad
phi matrix}) and~(\ref{A contravariant}) and the identity 
\begin{equation}
\nabla \cdot (c \vv{v}) = \nabla c \cdot \vv{v} + c \nabla \cdot \vv{v},
\end{equation}
we can write the Laplacian operator as 
\begin{eqnarray} \label{poisson cont expansion} \nonumber
\frac{1}{f}\nabla \cdot f \nabla \Phi &=& \nabla \cdot \vv{A} \\
\nonumber &=& \frac{1}{f}\nabla \cdot \left[f A^1 (\nabla \eta
  \times \hat{z}) + f A^2 (\hat{z} \times \nabla \xi) \right] \\
\nonumber &=& \frac{1}{f}\left[\nabla (f A^1) \cdot (\nabla \eta
  \times \hat{z}) + \nabla (f A^2) \cdot (\hat{z} \times \nabla
  \xi)\right].
\end{eqnarray} 
Using $\nabla \xi \cdot (\nabla \xi \times \hat{z}) = 0$,
(\ref{poisson cont expansion}) can be written as
\begin{equation}\label{nabla f1 and f2}
\frac{1}{f}\nabla \cdot f \nabla \Phi = 
\left[\frac{1}{f} \frac{\ptl (f A^1)}{\ptl \xi} \nabla \xi
  \cdot (\nabla \eta \times \hat {z})+\frac{1}{f} 
\frac{\ptl (f A^2)}{\ptl \eta} \nabla \eta \cdot (\hat{z}
  \times \nabla \xi)\right]. 
\end{equation} 
By (\ref{J from cross product}), we can rewrite (\ref{nabla
f1 and f2}) as
\begin{equation}
\frac{1}{f} \nabla \cdot f \nabla \Phi = \frac{1}{fJ} \left(\frac{\ptl
(fA^1)}{\ptl \xi}+\frac{\ptl (fA^2)}{\ptl \eta} \right)=-
\frac{\rho}{\epsilon_0}.
\end{equation}
Now converting the contravariant components of $\vv{A}$ to their
covariant formulations using \twoeqns{grad phi matrix}{A cov to cont
conv}, we can write the final form of the Poisson equation in
curvilinear coordinates:
\begin{equation}\label{curv coords poisson} 
\frac{1}{fJ} \frac{\ptl}{\ptl \xi^\alpha } \left(fJ g^{\alpha
\beta}\frac{\ptl \Phi}{\ptl \xi^\beta} \right) =-4\pi\rho,  
\end{equation} 
where $\rho = \rho^x$ is the physical charge density.

%%%%%%%%%%%%%%%%%%%%%%%%%%%%%%%%%%%%%%%%%%%%%%%%%%%%%%%%%%%%%%%%%%%%%%%
%%%%%%%%%%%%%%%%%%%%%%%%%%%%%%%%%%%%%%%%%%%%%%%%%%%%%%%%%%%%%%%%%%%%%%%
\section*{References}
\bibliographystyle{unsrt} \bibliography{CSDpaper}               %%%% REFERENCES %%%%
%%%%%%%%%%%%%%%%%%%%%%%%%%%%%%%%%%%%%%%%%%%%%%%%%%%%%%%%%%%%%%%%%%%%%%%
%%%%%%%%%%%%%%%%%%%%%%%%%%%%%%%%%%%%%%%%%%%%%%%%%%%%%%%%%%%%%%%%%%%%%%%
\end{document}